\documentclass[a4paper,12pt]{article}
\pdfoutput=1
\usepackage{jheppub}
\usepackage{amsmath,amssymb,epsf}
\usepackage{graphicx,color}

\usepackage{amssymb}
\usepackage{amsmath}
\usepackage{mathrsfs}
\usepackage{graphics}
\usepackage{epsfig}
\usepackage{graphicx}
\usepackage{subfigure}
\usepackage{bm}
\usepackage{nicefrac}
\usepackage{soul}

\def\to{\rightarrow}

\def\to{\rightarrow}

\def\6{\partial}

\newcommand{\bright}{\begin{flushright}}
\newcommand{\eright}{\end{flushright}}
\newcommand{\bminip}{\begin{minipage}}
\newcommand{\eminip}{\end{minipage}}
\newcommand{\bcent}{\begin{center}}
\newcommand{\ecent}{\end{center}}

\newcommand{\gsim}{\mbox{\raisebox{-.3em}{$\;\stackrel{>}{\sim}\;$}}}

\usepackage{amssymb,amsmath,euscript,latexsym}
\usepackage{jheppub}
\usepackage{graphicx}
\usepackage{hyperref}
\usepackage{color}

\def\dsl{\displaystyle}

\title{\bf{Supergravity Contributions to Inflation\\  \Large{in models with non-minimal coupling to gravity}}}

\author[a, c]{Kumar Das,}
\author[b]{Valerie Domcke,}
\author[a,c]{Koushik Dutta}
\affiliation[a] {Theory Division, Saha Institute of Nuclear Physics, 1/AF Saltlake, Kolkata, India}
\affiliation[b]{AstroParticule et Cosmologie (APC), Paris Centre for Cosmological Physics (PCCP), Universit\'e Paris Diderot, 75013 Paris, France}
\affiliation[c]{Homi Bhabha National Institute, Training School Complex, Anushaktinagar, Mumbai - 400094, India}
\emailAdd{kumar.das@saha.ac.in}
\emailAdd{valerie.domcke@apc.univ-paris7.fr}
\emailAdd{koushik.dutta@saha.ac.in}

\abstract{
This paper provides a systematic study of supergravity contributions relevant for inflationary model building in Jordan frame supergravity. In this framework, canonical kinetic terms in the Jordan frame result in the separation of the Jordan frame scalar potential into a tree-level term and a supergravity contribution which is potentially dangerous for sustaining inflation. We show that if the vacuum energy necessary for driving inflation originates dominantly from the F-term of an auxiliary field (i.e.\ not the inflaton), the supergravity corrections to the Jordan frame scalar potential are generically suppressed. Moreover, these supergravity contributions identically vanish if the superpotential vanishes along the inflationary trajectory. On the other hand, if the F-term associated with the inflaton dominates the vacuum energy, the supergravity contributions are generically comparable to the globally supersymmetric contributions. In addition, the non-minimal coupling to gravity inherent to Jordan frame supergravity significantly impacts the inflationary model depending on the size and sign of this coupling. We discuss the phenomenology of some representative inflationary models, and point out the relation to the recently much discussed cosmological
`attractor' models.}

\begin{document}
\maketitle
\section{Introduction}

The paradigm of cosmic inflation~\cite{Starobinsky:1980te,Guth:1980zm,Linde:1981mu} is by now well established in early universe cosmology, showing a remarkable success of the simplest single-field slow-roll inflation models when compared with the recent Planck data of the CMB fluctuations~\cite{Ade:2015lrj}. Typically, inflation is assumed to have occurred at some very high energy scale and in its simplest realization, a single scalar field is responsible for its dynamics. {Supergravity~\cite{Wess:1992cp}, arising also as the low energy limit of string compactifications, is a promising theoretical framework to describe inflation:} providing numerous (complex) scalar fields potentially suitable for inflation, it also consistently accounts for  the Planck-suppressed corrections to global supersymmetry, which can no longer be simply neglected at the high energy scales of inflation.

 On the one hand, these supergravity contributions represent a challenge for inflationary model building, potentially spoiling the flatness of the scalar potential required for slow-roll inflation. The $\eta$-problem of F-term inflation (with minimal coupling to gravity) is a well-known example of this problem~\cite{Copeland:1994vg,Dine:1995uk}. On the other hand, inflation models with a non-minimal coupling to gravity have recently received a lot of interest, e.g.\ in the context of Higgs inflation~\cite{Bezrukov:2007ep} and so-called attractor models~\cite{Kallosh:2013yoa}. A key observation is that including the non-minimal coupling to gravity, many inflation models with very different scalar potentials asymptotically approach the same unique predictions for the spectral index $n_s$ and the tensor to scalar ratio $r$, which moreover lie right in the sweet spot of the recent Planck data \cite{Einhorn:2009bh,Ferrara:2010yw,Buchmuller:2013zfa,Giudice:2014toa,Pallis:2013yda,Pallis:2014dma,Pallis:2014boa,Ellis:2013xoa,Kallosh:2013xya,
Nakayama:2010ga}. Such a non-minimal coupling to gravity is a characteristic feature of supergravity in the Jordan frame~\cite{Kallosh:2013hoa,Kallosh:2013tua,Ferrara:2013rsa,Kallosh:2013yoa,Galante:2014ifa,Broy:2015qna}.

In this work we systematically classify the supergravity contributions to Jordan frame inflation models. Starting from the framework of canonical superconformal supergravity (CSS) models suggested in \cite{Ferrara:2010in}, characterized by canonical kinetic terms in the Jordan frame and an approximate conformal symmetry, we determine the generic properties of supergravity contributions to the Jordan frame scalar potential, as well as the resulting contributions to the Einstein frame Lagrangian. This enables us to disentangle two important supergravity effects: contributions to the (Jordan frame) scalar potential which typically come in powers of $\phi/M_{pl}$ yielding dangerous corrections at large field values and contributions to the kinetic term in the Einstein frame, which in many cases lead to a flattening of the potential, favourable for slow-roll inflation. 

Our analysis reveals that the CSS models provide a powerful model building framework to control supergravity contributions. If the vacuum energy driving inflation is dominated by the F-term associated with the chiral multiplet of the inflaton, {the supergravity contributions become comparable to the globally supersymmetric contributions but do not necessarily dominate at large field values.} If the F-term of the inflaton field is subdominant, {the dangerous supergravity contributions to the scalar potential are generically suppressed at large field values compared to the contributions from global supersymmetry}. We further investigate under which conditions these generic supergravity contributions to the (Jordan frame) scalar potential vanish, finding (for single-field inflation) that this can only be achieved if the F-term of the inflaton is subdominant and if the superpotential vanishes along the inflationary trajectory. Turning to the kinetic term in the Einstein frame, we generalize the results obtained in the context of $\alpha$-attractors~\cite{Galante:2014ifa}, demonstrating that non-canonical kinetic terms lead to an exponential flattening of the potential if the functional dependence of the Jordan frame scalar potential and the non-minimal coupling to the Ricci scalar on the inflaton field are adjusted accordingly. {The combination of the (mildly broken) conformal symmetry inherent to CSS models in combination with specific choices of superpotentials allows for inflation models in agreement with current observations.}

This paper is organized as follows. After a brief review of supergravity in the Jordan frame in Sec.~\ref{review}, introducing in particular the framework of CSS models, we turn to the supergravity contributions to the Jordan frame scalar potential in Sec.~\ref{sec:sugra_contributions}. In Sec.~\ref{sec:criteria} we derive the conditions under which these generic contributions vanish. We demonstrate that for (effective) single-field inflation these conditions are equivalent to the requirement of a vanishing superpotential along the inflationary trajectory, and illustrate these results by applying them to some well-known inflation models: Monomial inflation, hybrid inflation and tribrid inflation. The details of the latter are left to App.~\ref{appendixa}. In Sec.~\ref{sec:attractors} we turn to supergravity contributions in the kinetic term in the Einstein frame, shedding light on the feature of `attractor' models in this set-up. Sec.~\ref{sec:examples} discusses two explicit examples in which all the above effects are illustrated. We conclude in Sec.~\ref{conclusion}.

\section{Supergravity in the Jordan frame \label{review}}

A supergravity model in $\mathcal{N}=1$ and $d=4$ is most commonly described in the Einstein frame where {the} matter part is minimally coupled to gravity, and the scalar and gravitational parts of the Lagrangian density relevant for inflation are determined by the superpotential $W$ and the K\"ahler potential $K$~\cite{Wess:1992cp}
\begin{align} \label{Lagrangian_Einstein_frame}
{\cal L}_E^\text{grav} + {\cal L}_E^\text{scal} = \sqrt{-g_E} \left( \frac{1}{2} {\cal R}(g_E) M_{pl}^2 +  g_E^{\mu \nu} K_{\alpha \bar \beta} (\partial_\mu z^\alpha) (\partial_\nu \bar z^{\bar \beta}) - V_E \right)\,,
\end{align}
with the metric $g^{\mu \nu}_{E}$, the Ricci scalar ${\cal R}$ and the field-space metric $K_{\alpha \bar \beta} = \partial_\alpha \partial_{\bar \beta} K$, obtained by taking the respective derivatives of $K$ with respect to the matter fields $z^\alpha$ and their conjugates $\bar z^{\bar \alpha}$, $\alpha = \{1,2,\dots \}$. $V_E$ denotes the scalar potential, and for the purpose of this paper we will focus on the F-term contribution,
\begin{align} \label{f_term_einstein}
V_E^F = e^{K/M_{pl}^2} \bigg[K^{\alpha\bar{\beta}}{\cal D}_{\alpha}W {\cal D}_{\bar{\beta}}\bar{W} - \frac{3 |W|^2}{M_{pl}^2} \bigg]. 
\end{align}
Here ${\cal D}_{\alpha}W = W_\alpha + W K_\alpha/M_{pl}^2$ with the subscripts denoting partial derivatives with respect to $z^\alpha$, $\partial_\alpha Y = Y_\alpha$.  $K^{\alpha\bar{\beta}}$ is the inverse of K\"ahler metric $K_{\alpha \bar \beta}$. The exponential factor in Eq.~(\ref{f_term_einstein}) is the source of the usual $\eta$-problem in supergravity inflation, which needs to be overcome when constructing inflation models in supergravity~\cite{Copeland:1994vg,Dine:1995uk}. {This may be achieved by either imposing a symmetry in the K\"ahler potential~\cite{Kawasaki:2000yn, Antusch:2008pn}, or tuning the model parameters \cite{Dvali:1994ms}.\footnote{As an alternative solution to $\eta$-problem, please see~\cite{Germani:2010gm}.} Additional contributions to the $\eta$-parameter arise from the other terms of the $F$-term scalar potential, and these typically depend on the form of the superpotential during inflation.} 

On the other hand, a $\mathcal{N} = 1$, $d= 4$ supergravity model may also be considered in a Jordan frame characterized by a frame function $\Phi(z^\alpha, \bar z^{\bar \alpha})$. In this case, the gravitational and the scalar parts of the Lagrangian {density} read~\cite{Kallosh:2000ve, Ferrara:2010yw, Ferrara:2010in}
\begin{align} \label{L_gravity}
\mathcal{L}_{J}^{grav}& = -\sqrt{-g_{J}} \, \frac 1 6 \, \Phi(z,\bar{z}) \, R (g_{J}) \,,\\ 
\mathcal{L}_{J}^{scalar}&= \sqrt{-g_{J}}\bigg[ \bigg(\frac{\Phi K_{\alpha\bar{\beta}}}{3 M_{pl}^2}- \frac{\Phi_{\alpha}\Phi_{\bar{\beta}}}{\Phi}\bigg)g_{J}^{\mu\nu}(\partial_{\mu}z^{\alpha})(\partial_{\nu}\bar{z}^{\bar{\beta}})-V_J \bigg]\,.  \label{scalar_actn}
\end{align}
 We note from Eq.~\eqref{L_gravity} that the matter fields $z^{\alpha}$ are non-minimally coupled to gravity.  A conformal transformation allows us to switch from a Jordan frame to the Einstein frame Lagrangian of Eq.~\eqref{Lagrangian_Einstein_frame}:
\begin{align} 
g_{\mu\nu}^J =\Omega^2\,  g_{\mu\nu}^E \quad \text{where,} ~~~~~ \Omega^2 = -\frac{3 M_{pl}^2}{\Phi}> 0 \,.
\end{align}
In particular, the Jordan frame scalar potential is related to the scalar potential in the Einstein frame as
\begin{align} \label{Jordon_to_Einstein}
V_J = \frac {\Phi^2}{9 M_{pl}^4}V_{E}  \,.
\end{align}
If $\Phi = -3 M_{pl}^{{2}}$ the transformation is trivial; this particular Jordan frame represents the Einstein frame. 

Non-trivial Jordan frame inflation models are characterized by a non-minimal coupling of gravity to the inflaton field, a feature which has recently received a lot of interest in the context of Higgs inflation~\cite{Bezrukov:2007ep,Einhorn:2009bh,Ferrara:2010yw}. Moreover, the Jordan frame has intriguing properties from a more conceptual point of view. As was demonstrated in Ref.~\cite{Kallosh:2000ve}, the usual formulation of supergravity can be obtained by starting from a larger symmetry group, namely the superconformal group, and gauge-fixing the additional degrees of freedom. This approach naturally leads to Jordan frame supergravity models, which can then be translated to the Einstein frame by a conformal transformation. {All inflationary observable quantities are frame independent, and can be calculated in either the Jordan or the Einstein frame~\cite{George:2013iia,Chiba:2008ia}.} However,  the simplicity of a given model may be obscured depending on the frame used.
 
A class of models inspired by this approach are the canonical superconformal  supergravity (CSS) models, cf.\ Ref.~\cite{Ferrara:2010in}, which feature an intriguingly simple structure in the Jordan frame. They are characterized by {the choice}
\begin{align}
\Phi(z,\bar{z})&= |z^0|^2 +  \delta_{\alpha\bar{\beta}} z^{\alpha} \bar{z}^{\bar{\beta}}  \mapsto -3M_{pl}^2 + \delta_{\alpha\bar{\beta}} z^{\alpha} \bar{z}^{\bar{\beta}} \,,
\label{frame}      \\
K(z,\bar{z})&=-3M_{pl}^2\ln{\bigg(-\frac{\Phi(z,\bar{z})}{3M_{pl}^2}\bigg)} \,,
\label{kahler}
\end{align}
with $z_0 \mapsto \sqrt{3} M_{pl}$ denoting the gauge-fixing of the conformal compensator field.
Eqs.~\eqref{frame} and \eqref{kahler} lead to
\begin{equation} \label{Kahler_metric}
K_{\alpha\bar{\beta}} = -\frac{3 M_{pl}^2}{\Phi} \bigg( \delta_{\alpha\bar{\beta}} - \frac{\Phi_{\alpha}\Phi_{\bar{\beta}}}{\Phi} \bigg) \,,
\end{equation}
and hence {Eqs.~\eqref{L_gravity} and \eqref{scalar_actn} become}
\begin{align}
\frac{1}{\sqrt{-g_J}}\mathcal{L}_J^\text{grav + scal} = \underbrace{\frac 1 2 \, M_{pl}^2\,  R (g_J)}_{\text{pure ~ SUGRA}} - \underbrace{\frac 1 6 \, M_{pl}^2 \, R (g_J) \, |z|^2 - \delta_{\alpha\bar{\beta}} \, g_{J}^{\mu\nu}(\partial_{\mu}z^{\alpha})(\partial_{\nu}\bar{z}^{\bar{\beta}}) -  V_J}_{\text{superconformal ~ matter}}\,.
\label{eq:CSSlagrangian}
\end{align} 
This Jordan frame Lagrangian shows several remarkable features. For a suitable (scale-invariant) scalar potential, the matter sector features a superconformal symmetry. This in particular includes invariance under local conformal transformations,
\begin{equation}
g_{\mu \nu} \rightarrow e^{- 2 \alpha(x)} g_{\mu \nu} \,, \quad z \rightarrow e^{\alpha(x)} z \,, \quad \bar z \rightarrow e^{\alpha(x)} \bar z \,.
\end{equation}
The first term in Eq.~\eqref{eq:CSSlagrangian} breaks this symmetry \cite{Moon:2009zq}, which can be traced back to the fixing of the conformal compensator.
The kinetic terms of the matter fields $z^\alpha$ are canonical\footnote{In general, there are Planck-suppressed corrections to this, proportional to the bosonic part of the auxiliary field of the supergravity Weyl multiplet, these however vanish on inflationary trajectories along the purely real or imaginary part of any single $z^\alpha$.} and the F-term scalar potential can be expressed as~\cite{Einhorn:2012ih,Buchmuller:2012ex}
\begin{align}
V_{J}^F & = V^\text{glob}  + \Delta V_J \\
 & = \delta^{\alpha\bar{\beta}} \, W_{\alpha}\bar{W}_{\bar{\beta}} + \frac{1}{\Delta_K }|\delta^{\alpha\bar{\beta}}\,W_{\alpha}\Phi_{\bar{\beta}} -3 W|^2 
\label{potn_jr}  
\end{align}
where, 
\begin{align}
\qquad \Delta_K  = \Phi - \delta^{\alpha\bar{\beta}}\,\Phi_{\alpha}\Phi_{\bar{\beta}} \,.
\label{eq_Delta}
\end{align}
Here the first term in Eq.~(\ref{potn_jr}) corresponds to the scalar potential obtained in global supersymmetry and the second term represents the supergravity contributions in the Jordan frame. 

Couplings between the conformal compensator and the matter fields of the theory lead to additional operators in the frame function which break the conformal symmetry. Following Ref.~\cite{Ferrara:2010in}, we will consider the following two operators,
\begin{equation}
\delta \Phi =  \chi \left(a_{ab} \frac{z^a z^b \bar z^{\bar 0}}{z^0} + h.c. \right) - 3 \zeta \frac{|z^n \bar z^{\bar n}|^2}{z^0 \bar z^{\bar 0}} \,,
\label{eq:deltaphi}
\end{equation}
with $\{a,b\}$ and $n$ running over distinct subsets of $\{1,2,...\}$. In particular, we will be interested in the case where the first term provides an additional parameter to the potential of the inflaton (which we will refer to as $\phi$ in following), $a_{ab} = \delta_{a \phi}\delta_{b \phi}/2$, whereas the second term {(with $\zeta$ being positive)} may stabilize orthogonal directions to inflaton in the field space~\cite{Kawasaki:2000yn, Ferrara:2010in}. We will in particular be interested in employing this term for so-called stabilizer fields, commonly denoted by $X$. After gauge-fixing the conformal compensator, the frame function then finally reads
\begin{align}
\Phi(z,\bar{z})= -3 M^2_{pl} + |z^{\alpha}|^2  + \frac{\chi}{2}(\phi^2 + \bar \phi^2) - \zeta|X \bar{X} |^2/M_{pl}^2\,.
\label{framefct}
\end{align}
Note that the first term in Eq.~\eqref{eq:deltaphi} does not modify Eq.~\eqref{Kahler_metric} (canonical kinetic terms in the Jordan frame) and Eq.~\eqref{potn_jr}, since this holds for all frame functions with $\Phi_{\alpha \bar \beta} = \delta_{\alpha \bar \beta}$.{} The second term in Eq.~\eqref{eq:deltaphi} on the other hand modifies Eq.~\eqref{potn_jr}, leading to 
\begin{align}
V_{J}^F = \delta^{\alpha\bar\rho}\delta^{\gamma\bar{\beta}}\Phi_{\alpha\bar{\beta}} \, W_{\gamma} \, \bar{W}_{\bar{\rho}} + V'_{\Delta\text{sugra}} \,,
\label{eq_potn_jr_2}
\end{align}
with $V'_{\Delta \text{sugra}}$ denoting a lengthy expression, whose explicit form is little enlightening. In the limit $X \rightarrow 0$, Eq.~\eqref{eq_potn_jr_2} however reduces to Eq.~\eqref{potn_jr} with $V'_{\Delta\text{sugra}}\stackrel{X\to 0}\longmapsto V_{\Delta\text{sugra}}$. This may be verified by considering the K\"ahler metric
\begin{equation}
K_{\alpha\bar{\beta}} = -\frac{3 M_{pl}^2}{\Phi} \bigg( \Phi_{\alpha\bar{\beta}} - \frac{\Phi_{\alpha}\Phi_{\bar{\beta}}}{\Phi} \bigg)\,,
\end{equation}
and the K\"ahler derivatives
\begin{equation}
{\cal D}_\alpha W = W_\alpha - 3 W \frac{\Phi_{\alpha}}{\Phi} \,,
\end{equation}
confirming that all contributions stemming from the second term in Eq.~\eqref{eq:deltaphi} vanish when the $X$-field is successfully stabilized at zero vev along the inflationary trajectory. The same conclusion holds for the corrections to the canonical kinetic terms in the Jordan frame. Consequently, Eq.~\eqref{potn_jr} will prove to be a powerful guide to find models unspoiled by supergravity corrections, even when including higher-dimensional operators in the frame function which break the conformal symmetry. In the following we will 
work in units of the reduced Planck mass, $M_{pl}=1$.

\section{Supergravity corrections to the scalar potential}
\label{sec:sugra_contributions}

{In the previous section, we noted that with the choice of the K\"ahler potential of Eq.~\eqref{kahler}, where the frame function $\Phi$ is given by Eq.~\eqref{framefct}, there is a clear separation between the globally supersymmetric and the supergravity contribution to the scalar potential in the Jordan frame.} In this section we determine the importance of this supergravity contribution in generic inflation models. In particular we will focus on single field inflation models, in which {$\chi < 0$}, the inflaton direction is given by $\varphi = {\sqrt{2}}\text{Re}(\phi)$ and all stabilizer fields $X_i$ are stabilized at zero. See Sec.~\ref{sec:examples} for explicit examples of this type.
Introducing
\begin{equation}
 f(\varphi) \equiv \langle \delta^{\alpha\bar{\beta}} \, 
W_{\alpha}  \Phi_{\bar{\beta}}  - 3W \rangle \,, \quad g^2(\varphi) \equiv V^\text{glob} \,,
\label{eq:fphi}
\end{equation}
where the expectation value $\langle .. \rangle$ indicates that the corresponding terms are to be evaluated along the inflationary trajectory, the Jordan frame potential (with the frame function~\eqref{framefct}) for the inflaton reads
\begin{align}
V_J(\varphi) & = V^\text{glob} + \Delta V_J  \label{eq:DeltaVJa}\\
 & =  g^2(\varphi) - \frac{|f(\varphi)|^2/3}{1 + \chi \, (1 + \chi) \, \varphi^2  /6} \,,
\end{align}
where we have employed the frame function~\eqref{framefct}. {We recall that in the Jordan frame the kinetic term of the scalar field $\varphi$ is canonically normalized.} Let us assume that $f(\varphi)$ and $g(\varphi)$ can be expressed as polynomials of order $p_f$ and $p_g$ in $\varphi$, respectively,
\begin{equation}
g(\varphi) = \sum_{n = 1}^{p_g} a_n \varphi^n \,, \qquad |f(\varphi)| = \sum_{n = 1}^{p_f} b_n \varphi^n \,,
\end{equation}
with generic coefficients $\{a_n, b_n\} = {\cal O}(1)$.
In the large field limit $\varphi^2 |\chi (1 + \chi)|/6 \gg 1$, $|\varphi| \gg 1$, {the above potential} simplifies to
\begin{equation}
V_J(\varphi) \simeq (a_{p_g} \varphi^{p_g})^2 - \frac{2 \, (b_{p_f} \varphi^{p_f})^2}{\chi \, (1 + \chi) \, \varphi^2 } \,.
\label{eq:pgpf}
\end{equation}
The supergravity correction term {appears} with a suppression factor of $1/(\varphi^2 \chi (1 + \chi))$, which will suppress  the supergravity corrections iff
\begin{equation}
p_f \leq p_g \,,
\end{equation}
i.e.\ if the maximal power of $\varphi$ in $f(\varphi)$ is not larger than the maximal power of $\varphi$ in $g(\varphi)$. 

What is the relation between $p_f$ and $p_g$ in generic inflation models? To answer this question, let us first make some simplifying assumptions: (i) we will consider only single-field inflation models, i.e.\ models in which the inflaton field $\varphi$ is the only dynamical field. (ii) all other fields are hence stabilized at fixed value in field-space during inflation, without loss of generality we can take this value to be zero. We will denote these fields by $X_i$. 
We will consider two limiting cases, $|F_\phi| \gg \sum_i |F_{X^i}|^2$ and $|F_\phi| \ll \sum_i |F_{X^i}|^2$. 

In the first case, $V^\text{glob} \simeq |F_\phi|^2 \sim |W/\phi|^2$.\footnote{{The latter relation assumes that the term in the superpotential responsible for the dominant contribution to the vacuum energy also yields the dominant contribution to $W$. In the limit of large $\phi$, this assumption is justified.}} With $\Phi$ as in Eq.~\eqref{framefct}, the power $p_f$ of $f(\varphi)$ is given by the power of $\phi$ in $W(\phi)$  - in the absence of cancellations among the terms in Eq.~\eqref{eq:fphi}. We will return in detail to the possibility of such cancellations in {the next section.} For now we assume that they are absent, leading to $p_f = p_g + 1$. Thus, for large values of $\varphi$, the second term in Eq.~\eqref{eq:pgpf} is generically of the same size as the globally supersymmetric term.

On the other hand, for $|F_\phi| \ll \sum_i |F_{X^i}|^2$, {the globally supersymmetric scalar potential is given by  $V_J^\text{glob} \simeq \sum_i|F_{X^i}|^2 $, whereas the supergravity contributions are controlled (in the large field limit) by
\begin{equation}
\Delta V_J = -\frac{2 \, |f(\varphi)|^2}{\chi\, (1 + \chi)\, \varphi^2 } \sim \frac{\langle |W| \rangle ^2}{\varphi^2} \sim |F_\phi|^2 \ll  \sum_i |F_{X^i}|^2 \simeq V_J^\text{glob}\,,
\end{equation}
where we have used $\langle W \rangle \sim F_\phi \phi$ since the terms in $W$ proportional to $X^i$ vanish along the inflationary trajectory. We have also dropped factors of $\chi$ and $(\chi + 1)$, assuming that in the large field limit, $\chi (1 + \chi) \varphi^2/6 \gg 1$, these are roughly order one factors. With this, we conclude that for $|F_\phi| \ll |F_{X^i}|$, the supergravity contributions to the Jordan frame scalar potential are always suppressed.}

To illustrate this point, consider e.g.\ the superpotential
\begin{equation}
W = \lambda X \phi^2 + \epsilon \phi^{n}  \,,
\label{eq:Weps}
\end{equation}
where we assume for the moment $\langle X \rangle = 0$ and $\langle \text{Im}(\phi) \rangle = 0$, with the inflaton given by  $\varphi =  \sqrt{2}~ \text{Re}(\phi)$. The $F$-terms are give by $\langle F_X \rangle = \lambda \langle \phi^2 \rangle $ and $\langle F_\phi \rangle = n \epsilon \langle \phi^{n-1} \rangle$, respectively.
The first term in Eq.~\eqref{eq:Weps} is the usual superpotential of monomial inflation, which we will return to in Sec.~\ref{sec:examples}. 
The second term adds a non-vanishing $F_\phi$-term, tunable with the parameter $\epsilon$. With the K\"ahler potential~\eqref{kahler} and the frame function~\eqref{framefct}, the scalar potential for the inflaton in the Jordan frame reads
\begin{align}
{V_J = \frac{1}{2} \lambda^2 \varphi^4 + \epsilon^2 \frac{n^2}{2^{n - 1}} \varphi^{2n - 2} - \epsilon^2 \frac{((n- 3 )  + \chi n )^2}{2^n \left[1 + \varphi^2 \chi (1 + \chi)/6 \right]} \varphi^{2n}}  \,,
\label{eq:Veps}
\end{align}
The first two terms here correspond to the bare potential {obtained in global supersymmetry} and the last term stems from the supergravity contribution. If $|F_\phi| \gg |F_X|$, the globally supersymmetric scalar potential is dominated by the second term (proportional to $\epsilon^2 \varphi^{2n - 2}$), and is thus of the same order as the supergravity contribution in the large field limit. On the other hand, if $|F_\phi| \ll |F_X|$, the first term will dominate the scalar potential and the supergravity contribution is suppressed. Note that for $n > 3$, the condition $|F_\phi| \ll |F_X|$ will be violated at very large field values, bringing us back to the scenario discussed above. We will return to this toy-model at the end of Sec.~\ref{sec:attractors}, where we will in particular focus on the case $n = 3$. 

The considerations above focused on large field values for $\varphi$. Of course, for small field values, the supergravity contributions become parametrically suppressed by $\varphi/M_{pl}$. The question of supergravity contributions is thus a question of large field inflation models.

In summary, we conclude that by construction, the CSS framework prevents excessive supergravity contributions to the Jordan frame scalar potential. The supergravity contributions can be at most of the same power in the inflaton field as the globally supersymmetric contribution, and in inflation models whose vacuum energy is generated by F-terms of fields other than the inflation, they are even subdominant compared to the globally supersymmetric contribution in the large field limit.

\section{Criteria for vanishing supergravity contributions in \texorpdfstring{$V_J$}{VJ}}
\label{sec:criteria}
In the previous section we discussed the generic size of the supergravity contributions to the scalar potential in the Jordan frame, {and for that purpose, we did not assume any specific form of the superpotential.} In this section we investigate under which conditions the above generic conclusions can be evaded, more specifically under which conditions the supergravity contribution to the Jordan frame scalar potential vanishes. This will bring us to the questions of a possible cancellation in Eq.~\eqref{eq:fphi}, as mentioned {in the previous section.}

Let us start with writing the general form of the superpotential $W(z)$ as
\begin{align}
W&=W^{(0)} + W^{(1)} + W^{(2)} + W^{(3)} + W^{(4)} + \dots \,,
\label{wgeneral}
\end{align}
with the superscript $i = \{1,2,..\}$ denoting the number of superfields in the respective term, i.e.
\begin{align}
W= W^{(0)} + a^1_\alpha z^\alpha + a^2_{\alpha \beta} z^\alpha z^\beta +a^3_{\alpha \beta \gamma} z^\alpha z^\beta z^\gamma + a^4_{\alpha \beta \gamma \delta} z^\alpha z^\beta z^\gamma z^\delta + \dots \,,
\end{align}
with $W^{(0)}$ and $a^i$ are constants of the theory. Denoting the inflaton field as $\phi$, the frame function of Eq.~\eqref{framefct} reads 
\begin{align}
\Phi = -3 + |z^{\alpha}|^2 + \frac{\chi}{2}(\phi^2 +\bar{\phi}^2) - \zeta|X^i|^4 ~,
\label{eq:frame_simple}
\end{align}
with $\alpha = \{1,2,..\}$ running over all the fields of the theory, and $X^i$ denoting a subset of fields not containing the inflaton field. 

The second term of Eq.~(\ref{potn_jr}) containing the supergravity contributions vanishes if
\begin{align}
|\delta^{\alpha \bar{\beta}} \, W_\alpha  \Phi_{\bar{\beta}}  -3 W|^2 = 0 \,,
\end{align}
which along the inflationary trajectory can be rewritten using Eq.~\eqref{wgeneral} as
\begin{align}
\langle W^{(4)} - W^{(2)} - 2W^{(1)} - 3W^{(0)} - 2\zeta \,  W_{X^i} X^i |X^i|^2 + \chi W_\phi \bar{\phi }\rangle = 0 \,.
\label{eq:condition}
\end{align}
Note that both $\chi$ and $\zeta$ are parameters of the K\"ahler potential. Barring fine tuning between the parameters of the K\"ahler potential and the parameters of the superpotential, Eq.~\eqref{eq:condition} can be expressed as
\begin{align}
\mathcal{C}:\, 
\begin{cases}
& \langle W^{(4)} - W^{(2)} - 2W^{(1)} - 3W^{(0)} \rangle = 0  \,, \\
& \langle \chi W_\phi \bar\phi - 2\zeta \, W_{X^i} X^{i} |X^i|^2 \rangle =0 \,.
\end{cases} 
\label{impt_condt}
\end{align}
Note the special role of the trilinear term in $W$, which, invariant under the conformal symmetry, is not constrained by the first part of this condition.
The conditions~\eqref{eq:condition} and \eqref{impt_condt} can easily be generalized if several fields receive holomorphic contributions in the frame function, i.e.\ for $ \chi/2(\phi^2 + \bar \phi^2) \mapsto \chi a_{ab}(z^a z^b + \bar z^a \bar z^b)$ in Eq.~\eqref{eq:frame_simple}. In this case the we find
\begin{equation}
\chi W_\phi \bar \phi \mapsto \chi  W_{a} (a_{ab} + a_{ba}) \,  \bar z^b 
\end{equation}
in Eqs.~\eqref{eq:condition} and \eqref{impt_condt}. The first line of Eq.~\eqref{impt_condt} remains unchanged. 

Now, a comment on the strength of these conditions is in order. The conditions ${\cal C}$ protect the direction of the (complex) inflaton field $\phi$ from large supergravity corrections, however other, orthogonal directions in the field space may still receive large (possibly tachyonic) masses, see eg.~\cite{Ferrara:2010yw,Ferrara:2010in,Lee:2010hj}. Tachyonic directions orthogonal to the inflationary trajectory render the latter unstable.
For `stabilizer'-type fields, which feature a vanishing vev during and after inflation, this may technically be remedied by adding the above mentioned $-\zeta |X^i|^4$ term to the K\"ahler potential. However for hybrid-inflation models which contain a `Higgs'-type field whose dynamics is responsible for ending inflation, the problem is more severe. In this case, a $-\zeta |X^i|^4$ term in the frame function stabilizing this field during inflation will generically also do so after inflation, thus preventing the desired phase transition ending inflation. 


\subsection{Connection to \texorpdfstring{$\langle W \rangle = 0$}{W=0} in single field inflation}
\label{wnonzero}

We now highlight how in single field inflation, under some {generic} conditions, the conditions for vanishing supergravity corrections~\eqref{impt_condt} are equivalent to the requirement of a vanishing superpotential along the inflationary trajectory, $\langle W \rangle = 0$. Considering the expression for the F-term scalar potential in the Einstein frame, Eq.~\eqref{f_term_einstein}, it is well known that models with $\langle W \rangle \neq 0$ obtain dangerous supergravity corrections due to the $- 3 |W|^2/M_{pl}^2$ term. Here we extend this understanding, showing how in the class of CSS models, $\langle W \rangle = 0$ is equivalent to the exact vanishing of the supergravity contribution in the Jordan frame scalar potential, $\Delta V_J = 0$.

Let us first consider an inflation model in which Eq.~\eqref{impt_condt} is fulfilled. Then for $\chi \neq 0$ the second line of Eq.~\eqref{impt_condt} requires $\partial W/\partial \phi = 0$ along the inflationary trajectory (assuming that all $X^i$'s are stabilized at zero {vev} by the $\zeta_i$-terms). Hence, assuming that {during} inflation there is only one\footnote{this includes models of hybrid-like inflation, where a second field becomes dynamical at the end of inflation.} dynamical field, {which is the inflaton} $\phi$, we can write
\begin{equation} \label{Wsimple}
 \langle W \rangle = \langle c_0 \rangle + \langle c_1 \rangle \phi + \langle c_2 \rangle \phi^2 + \dots ~.
\end{equation}
{When we require $\partial W/\partial \phi = 0$ along the inflationary trajectory over a finite range of $\phi$, the terms of different order in $\phi$ have to vanish independently, i.e  $\langle c_1 \rangle = \langle c_2 \rangle = \cdots = 0$.} We hence obtain $ \langle W \rangle  = \langle c_0 \rangle$, with $\langle c_0 \rangle$ being a function of the fields other than the inflaton.
Due to our assumption that the inflaton is the only dynamical field during inflation, all other fields which might enter {into} $\langle c_0 \rangle$ are characterized by a constant vev during inflation, and without loss of generality, we may set these vevs to zero.\footnote{More generally, we may allow any constant vev for these fields during inflation: Starting from $\langle X \rangle = c$, we define $X' = X - c$ and perform the above analysis for the redefined field $X'$.} 
Hence a non-zero $\langle c_0 \rangle$ can only be sourced by a true constant in the superpotential, $ W_\text{inf}  = W^{(0)} \neq 0$. This however is excluded by the first line of Eq.~\eqref{impt_condt} (again taking into account that this condition must hold over a finite range for $\phi$). In summary, under the assumption of (effectively) single field inflation, the condition~\eqref{impt_condt}, i.e.\ the vanishing of supergravity contributions to the Jordan frame scalar potential, implies $\langle W \rangle = 0$. {This is one of the main results discussed in this article.}

Conversely, under the assumption that the fields other than the inflaton are stabilized at constant vev during inflation, $\langle W \rangle =0$  guarantees that the conditions in Eq.~\eqref{impt_condt} are also satisfied. To prove this statement, we explicitly express the terms $W^{(i)}$ introduced in Eq.~\eqref{wgeneral} as a power expansion in $\phi$,
\begin{align}
W^{(i)}=\langle c_{i\,0}\rangle + \sum_{j=1}^{i} c_{ij} \phi^j = 
\langle c_{i\,0}\rangle + \langle c_{i\,1} \rangle \phi + \cdots + \langle c_{i\,i-1}\rangle\phi^{i-1} + c_{ii}\phi^i \,.
\end{align} 
Here $W^{(0)}$ and the $c_{ii}$'s are pure constant terms and $\langle c_{ij}\rangle$ for $i\ne j$ are the coefficients obtained from the vevs of all fields except inflaton $\phi$. Now if during inflation all fields other than the inflaton are stabilized at zero vev, then $\langle c_{\substack{ij \\ i \ne j}} \rangle = 0$. {We are} then {left with} $W = W^{(0)} +\sum_{i} c_{ii}\phi^i$. {Now, if we demand that} during inflation $W_\text{inf}=0$ is satisfied {for all inflaton field values}, this requires $W^{(0)}$ and $c_{ii}$ to be zero independently, and hence $W^{(0)} = W^{(1)} =W^{(2)} =W^{(3)} =W^{(4)} = ... = 0$. So the first part of the condition in Eq.~\eqref{impt_condt} is immediately satisfied. A little bit of more algebra shows that the second part of the condition too is fulfilled. In the context of usual Einstein frame supergravity, the advantage of $W_\text{inf} = 0$ for inflation model building has been noted earlier in \cite{Davis:2008fv, Antusch:2009ef, Mooij:2010cs}. 

In summary, the generic supergravity contributions to the Jordan frame scalar potential identified in Sec.~\ref{sec:sugra_contributions} can be avoided by a suitable construction of the superpotential. This leads in a first step to the condition of Eq.~\eqref{eq:condition}, which may be re-expressed as the two conditions of Eq.~\eqref{impt_condt} in the absence of correlations between the parameters of the K\"ahler and superpotential. For single-field inflation models, this further simplifies to the condition of a vanishing superpotential along the inflationary trajectory.
Returning to the two cases $|F_\phi| \ll \sum_i |F_{X^i}|$ and $|F_\phi| \gg \sum_i |F_{X^i}|$ discussed in Sec.~\ref{sec:sugra_contributions}, we note that the second line of Eq.~\eqref{impt_condt} immediately implies $F_\phi = \partial W/\partial \phi = 0$. Hence a cancellation of the supergravity contributions to the Jordan frame scalar potential by a suitable choice of superpotential is only possible for inflation models driven by a vacuum energy stemming from F-terms of fields other than the inflaton. Models with $|F_\phi| \gg \sum_i |F_{X^i}|$, which were found to generically obtain larger supergravity contributions in Sec.~\ref{sec:sugra_contributions}, cannot be protected in this way.

\subsection{Illustrative Examples}
\label{rep_models}

Let us apply the above mentioned derived condition for vanishing supergravity contributions in the Jordan frame to some well-known classes of inflation models. 

\subsubsection*{Hybrid Inflation} 
The superpotential of F-term hybrid inflation~\cite{Linde:1993cn,Dvali:1994ms,Copeland:1994vg} is given by 
\begin{equation}
W = \lambda \phi (H_+ H_- - M^2) \,,
\label{hybrid}
\end{equation}
with coupling constant $\lambda$, mass parameter $M$ and $H_\pm$ denote so-called waterfall fields which obtain non-zero vevs in a phase transition ending inflation. During inflation, $\langle H_\pm \rangle = 0$, and hence $\langle W \rangle \neq 0$. The resulting supergravity contributions induce a large tachyonic mass to the imaginary part of $\phi$, thus spoiling inflation~\cite{Buchmuller:2012ex}.

\subsubsection*{Monomial Inflation}
Monomial chaotic inflation is characterized by the following choice of the superpotential \cite{Linde:1983gd, Kawasaki:2000yn, Kallosh:2010xz}, 
\begin{align}
W= m X f(\phi) \,.
\end{align}
Here $\phi$ is the inflaton field and $X$ is an auxiliary field, often called a stabilizer field, which has a vanishing {vacuum expectation value} (vev) during inflation. Hence $W_\text{inf} = 0$ and the supergravity contributions vanish in the Jordan frame inflaton potential. We will return to this example in more detail in Sec.~\ref{sec:examples}.

\subsubsection*{Tribrid Inflation}
\label{tribrid_sub}
The  superpotential of tribrid inflation involves the interplay of three fields, i.e.\ the inflaton field $\phi$, the auxiliary field $X$ that is stabilized at zero vev, and the waterfall field $H$ which triggers the phase transition ending inflation~\cite{Antusch:2008pn, Antusch:2009vg, Antusch:2012bp, Antusch:2012jc, Antusch:2015tha}. The tribrid inflation superpotential can be defined by
\begin{align} \label{tribrid_eq}
W= \kappa X (H^l - M^2) + \lambda\phi^n H^m~,
\end{align}  
and for simplicity, we will take $l =2$. We will restrict ourselves to $n, m \leq 2$ in accordance with the maximum number of fields in $W$ given by Eq.~\eqref{wgeneral}. Here $\kappa,\lambda$ are {positive} constants. In contrast to the standard hybrid {inflation} case of Eq.~(\ref{hybrid}), there is an additional stabilizer field $X$ which drives inflation by providing a large vacuum energy through its $F$-term. During inflation $\langle X \rangle = \langle H \rangle =0$ and after inflation $ \langle H \rangle =M$. With $W_\text{inf}=0$ during inflation, the Jordan frame potential for the inflaton $V_J(\phi)$ is protected from supergravity contributions. 

However, the waterfall field $H$ is not protected in this way and may obtain large supergravity contributions. An indication of this can be obtained from evaluating Eq.~\eqref{impt_condt} slightly away from the inflationary trajectory, $H \rightarrow 0 + \epsilon$, in which case the first line of Eq.~\eqref{impt_condt} no longer vanishes. As mentioned in Sec.~\ref{sec:criteria}, Eq.~\eqref{impt_condt} does not guarantee the vanishing of supergravity corrections orthogonal to the inflaton trajectory - which is precisely the trouble we are running into here: the mass of the $H$ field is not protected from supergravity contributions. In Appendix.~\ref{appendixa} we explicitly calculate this contributions for $n = 2, m = 1, 2$, showing that for $n = 2, m = 1$ they can destabilize the inflationary trajectory whereas for $n = m = 2$, the model is more robust and could be a very interesting case for further study.

\section{Further supergravity effects \label{sec:attractors}}

In sections~\ref{sec:sugra_contributions} and \ref{sec:criteria}, we have discussed supergravity contributions to the Jordan frame scalar potential. However, a full analysis of the supergravity effects must also include the effects stemming from the non-minimal coupling to gravity (in the Jordan frame) or correspondingly from the conversion $V_E = \Omega^4 V_J$ and from the non-canonical kinetic terms (in the Einstein frame). An instructive analysis which directly applies to the {set-up} we discuss in the paper was given in Ref.~\cite{Galante:2014ifa} in the context of so-called universal $\xi$-attractors \cite{Kallosh:2013maa,Kallosh:2013tua,Galante:2014ifa}. These are characterized by a non-minimal coupling to the Ricci-scalar and the Lagrangian density is given by\footnote{{Note the slightly different notation compared to \cite{Galante:2014ifa}: $\Omega_\text{\cite{Galante:2014ifa}} = \Omega^{-2}$. For supergravity embeddings of this model see e.g.\ \cite{Kallosh:2010ug,Kallosh:2010xz,Einhorn:2009bh,Kallosh:2013hoa,Kallosh:2013tua}.}} 
\begin{equation}
\mathcal{L}_J = \sqrt{-g_J}\left[\frac{1}{2}\, \Omega^{-2}(\varphi)\,  R - \frac{1}{2} \, K_J(\varphi)\, (\partial\varphi)^2 - V_J(\varphi) \right]~,
\end{equation}
with $\Omega^{-2}(\varphi) = 1 + \xi \varphi^2$. For $K_J = 1$, $\xi = -(1 - |\chi|)/6$  this corresponds precisely to the setup discussed in this paper (once the real scalar inflaton field $\varphi$ has been identified), as will be illustrated in the explicit examples of Sec.~\ref{sec:examples}. 

With a suitable conformal transformation, the Lagrangian density above can be expressed in the Einstein frame, 
\begin{equation}
\label{eq:LE5}
\mathcal{L}_E = \sqrt{-g_E} \left[ \frac{1}{2} \mathcal{R}_E - \frac{1}{2} \underbrace{\left(\frac{K_J}{\Omega^{-2}} + 6 \, \frac{\Omega'^2}{\Omega^2}\right)}_{K_E(\varphi)} (\partial \varphi)^2 - \Omega^4 V_J(\varphi) \right] \,.
\end{equation}
 If ${K_J}{\Omega^2} \ll 6{\Omega'^2}/{\Omega^2}$ and $\xi > 0$ (i.e $|\chi| > 1$), the above Lagrangian can be re-casted (in terms of the dynamical variable {$\Omega$}) with a kinetic term having a second order pole at $\Omega = 0$ (corresponding to the large field limit $ \varphi \rightarrow \infty$)~\cite{Galante:2014ifa},
\begin{equation}
{\cal L}_E = \sqrt{- g_E} \left[\frac{1}{2} \mathcal{R}_E - 3 \left(\frac{\partial\Omega}{\Omega}\right)^2 - V_E(\Omega) \right]~.
\label{eq:LE2}
\end{equation}
From Eq.~\eqref{eq:LE2}, we see that the canonically normalized field $\hat \varphi$ is related to the dynamical variable $\Omega(\varphi)$ as $\Omega = \exp(- \hat \varphi/\sqrt{6})$.\footnote{We note that $\Omega(\varphi \rightarrow \pm \infty) = 0$. The canonical normalization of Eq.~\eqref{eq:LE2} leaves the sign ambiguity $\Omega = \exp(\pm \hat \varphi/\sqrt{6})$. We choose here the solution  $\Omega = \exp(- \hat \varphi/\sqrt{6})$, obtaining the desired asymptotic behaviour $\Omega(\hat \varphi \rightarrow + \infty) = 0$ for positive field values.}

Let us now consider a Jordan frame scalar potential $V_J(\varphi) = c_V \varphi^{2 l}$ (the situation is easily generalized to any polynomial scalar potential, whose maximum power of $\varphi$ is $2 l$). With $V_E = \Omega^4 V_J$ and $\varphi^2 = (\Omega^{-2} - 1)/\xi$ we find
\begin{align}
V_E &= c_V \xi^{-l} \Omega^4 \left( \Omega^{-2} - 1 \right)^{l} \\
& = c_V \xi^{-l} \left( \Omega^{4 - 2l} + \cdots +  \Omega^4 \right) \,.
\label{eq:VEOmega}
\end{align}
We can thus identify three qualitative different situations. (i) For $l < 2$, all powers of $\Omega$ appearing in Eq.~\eqref{eq:VEOmega} are positive and hence $V_E(\hat \varphi) \rightarrow 0$ for $\hat \varphi \rightarrow \infty$. Together with $V_E \simeq V_J \simeq c_V \hat \varphi^{2l}$ at small field values, this yields (for $c_V > 0$) a hilltop type potential (see e.g Fig.~\ref{pot_1}). (ii) For $l = 2$, the leading order term in $\Omega$ is constant, and hence these models approach an exponentially flat plateau for large field values (see eg Fig.~\ref{lphix_pot}). For $c_V > 0$ this reproduces the characteristic feature of the Starobinsky-type inflation models~\cite{Starobinsky:1980te}, which are favoured by the current CMB data. (iii) For $l > 2$, Eq.~\eqref{eq:VEOmega} contains negative powers of $\Omega$, leading to an exponential growth of the potential at large field values, unsuitable for slow-roll inflation. We will discuss explicit realizations of these different scenarios, in particular of the cases (i) and (ii), in the next section.

Note that the phenomenologically very promising case (ii) can more generally be obtained for any $V_J \propto f(\varphi)^2$ and $\Omega^{-2} = 1 + \xi f(\varphi)$, which has coined the name `attractor'-models~\cite{Bezrukov:2007ep,Kallosh:2013yoa,Buchmuller:2013zfa,Giudice:2014toa,Pallis:2013yda,Pallis:2014dma,Pallis:2014boa,Ellis:2013xoa,Kallosh:2013xya}, since different models are `attracted' to the sweet spot of the Planck data as the parameter $\xi$ is increased. These models might have very different potentials at small field values, but asymptotically they all feature an exponentially flat potential. The predictions in the $n_s$ - $r$ plane are described by a one-parameter region (here $\xi$) which converges to $(1-2/N, 12/N^2)$ at leading order in $1/N$  and for large $\xi$. The above analysis underlines that this mechanism requires the leading order power of $\varphi$ in $V_J^{1/2}$ and $\Omega^{-2}$ to be identical, which, as an ad hoc assumptions, requires some degree of tuning.

{We now discuss the case of $\xi < 0$ in $\Omega^{-2}(\varphi) = 1 + \xi \varphi^2$. For $0 < |\chi| < 1$, $\xi$ is bounded between $-1/6$ and $0$. In this case there is a pole in $\Omega$ for $\varphi \rightarrow 1/\sqrt{|\xi|}$. We thus make a change of variable to $\tilde \Omega = 1/\Omega$, and the Lagrangian of Eq.~\eqref{eq:LE5} becomes 
\begin{equation}
{\cal L}_E = \sqrt{- g_E} \left[\frac{1}{2} \mathcal{R}_E - 3 \left(\frac{\partial\tilde \Omega}{\tilde \Omega}\right)^2 - V_E(\tilde \Omega) \right]~.
\label{eq:LE3}
\end{equation}
In this case, the canonically normalized field $\hat \varphi$ is related to the dynamical variable $\tilde \Omega(\varphi)$ as $\tilde \Omega = \exp(- \hat \varphi/\sqrt{6})$.\footnote{{Note that if we consider negative field values, the corresponding canonically normalized field is given by $\tilde \Omega = \exp( \hat \varphi/\sqrt{6})$.}} Similar to the case for $\xi > 0$, the kinetic term now has a pole at $\tilde \Omega = 0$. As before, in terms of the canonically normalized field $\hat \varphi$, this pole is located at infinite field values. For $V_J = c_V \varphi^{2 l}$, substituting $\varphi$ by $\tilde \Omega$ yields
\begin{equation}
V_E= c_V \xi^{-l} \left( \tilde \Omega^{-4} + \cdots +  \tilde \Omega^{(2l - 4)} \right)  \,.
\label{eq:VEOmega1}
\end{equation}
For $l > 0$, we see that the first term always dominates the form the potential, and it is exponentially steep in terms of the canonical field $\hat \varphi$. Hence even if we get the required amount of inflation, it does not lead to the attractor prediction in the $n_s$-$r$ plane. We return to explicit examples of this type in the next subsection.}

Returning to our discussion of the Jordan frame scalar potential $V_J$ in Secs.~\ref{sec:sugra_contributions} and \ref{sec:criteria}, we note that the supergravity contribution always comes with a negative sign, i.e.\ $c_V < 0$ in the above parametrization. This implies that a viable inflation model will always require this contribution to be subdominant. This is easily achieved if the supergravity contribution comes with a power of $p_f < 3$ (see Eq.~\eqref{eq:pgpf}), corresponding to case (i) above. In this case the supergravity contribution will vanish in the large field limit. On the other hand, if $p_f > 3$ (case (iii)), the potential becomes unbounded from below at large field values. In the intermediate case (ii), the viability of the inflation model will depend on the relative size of the supergravity contribution, see example below.

\begin{figure}[t]
  \centering
   \includegraphics[width=8cm]{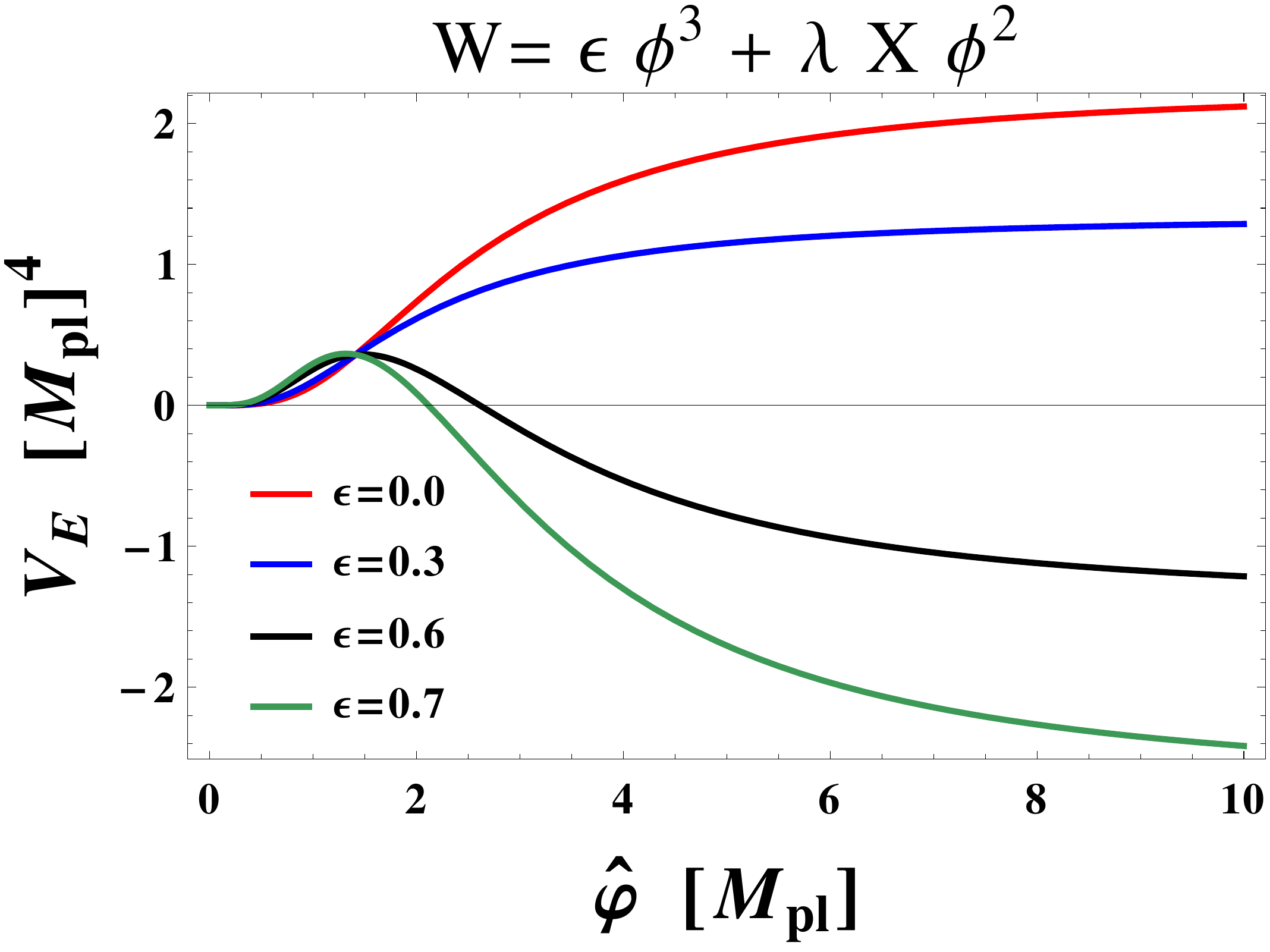} 
\caption{Behaviour of supergravity corrections for $W=\lambda X\phi^2 + \epsilon\phi^3 $. This plot shows the behaviour of the Einstein frame potential with incresing $\epsilon$ for $\lambda = 1$ . As we tune $\epsilon$ to higher values supergravity corrections become $\mathcal{O}(1)$.} 
\label{xyz}
\end{figure}

Let us illustrate these points by returning to the model introduced in Eq.~\eqref{eq:Weps} with $n = 3$. The scalar potential for the canonically normalized inflaton field in the Einstein frame is depicted in Fig.~\ref{xyz}. From Eq.~\eqref{eq:Veps}, we note that both the globally supersymmetric and the supergravity contribution come with a power of $\varphi^4$. For small values of $\epsilon$, the globally supersymmetric contribution dominates, leading to a positive plateau in the large field regime. For increasing values of $\epsilon$, the supergravity contribution becomes to dominate, and the scalar potential is found to take negative values in large field regime.

\section{Examples of inflation models}
\label{sec:examples}

To illustrate the analysis of the previous sections, we present two representative examples, based on $W = \lambda X \phi^2$ and $W = m X \phi$ as well as the frame function~\eqref{framefct},
\begin{align}
\Phi= -3 + \phi\bar{\phi} + X\bar{X} + \frac{\chi}{2}(\phi^2 + \bar{\phi}^2) -\zeta|X\bar{X}|^2  \,.
\label{fr_quadr}
\end{align}
Contrary to the example discussed at the end of the previous section, both examples here feature $\langle W \rangle = 0$ during inflation, hence the supergravity contributions to the Jordan frame scalar potential vanish and we are left with the type of supergravity effects described in Sec.~\ref{sec:attractors}. The phenomenology of these models has been previously studied in Ref.~\cite{Linde:2011nh}.

\subsection{Hilltop inflation from \texorpdfstring{$W=m\phi X$}{W= m phi} } \label{mphix}

For sufficiently large $\zeta$, the stabilizer field $X$ can be taken to be fixed at $\langle X \rangle = 0$, and hence following Eq.~\eqref{Jordon_to_Einstein}, the $F$-term scalar potential in Einstein frame is given by
\begin{align}
V_E|_{X\to 0} =\dsl{\frac{m^2(\varphi^2 + \tau^2)}{2\left(1-\frac{\varphi^2}{6}(1+\chi) + \frac{\tau^2}{6}(-1+\chi)\right)^2}}\,,
\label{VE_chaotic1}
\end{align}
where we have decomposed $\phi = (\varphi + i\, \tau)/\sqrt{2}$.
We first note that the resulting Lagrangian is invariant under the simultaneous transformation of $\varphi \leftrightarrow \tau$ and $\chi \rightarrow - \chi.$ We thus restrict our analysis to $\chi \leq 0$, for which the real part of $\phi$ will play the role of the inflaton.

In the Einstein frame, the kinetic term of the complex scalars $\phi$ and $X$ is not canonically normalized. We can however extract information about the dynamics of these fields, in particular the values of their masses and the inflationary slow-roll parameters, by exploiting
\begin{align}
\frac{d a}{d \hat a} &= \frac{1}{\sqrt{K_{z \bar z}}} \quad \text{for } \quad a = \text{Re}(z), \text{Im}(z) \\
&\rightarrow \quad \frac{\partial V}{\partial \hat a} = \frac{1}{\sqrt{K_{z \bar z}}} \frac{\partial V}{\partial a} \,,
\label{eq:derivatives}
\end{align}
where $\hat z$ (and correspondingly $\hat a$) denote the canonically normalized field.
With this, we verify that both $\tau$ (the imaginary part of the complex inflaton field $\phi$) and the complex stabilizer field $X$ are stabilized  at zero vev with $m^2_{\tau, X} \geq {\cal O}({\cal H}^2)$. For the stabilizer field $X$, this is achieved if $\zeta \gtrsim 0.7$ in the frame function. For the real scalar $\tau$, one may at first glance worry about about a pole in Eq.~\eqref{VE_chaotic1} for large values of $\tau$. However since the K\"ahler metric features the same pole structure, this pole is not reached for any finite value of the canonically normalized field $\hat \tau$.

We now turn in more detail to the inflationary dynamics. Along the inflationary trajectory, the scalar potential in the Einstein frame reduces to 
\begin{align}
V_E =\dsl{\frac{m^2 \varphi^2}{2\left(1-\frac{\varphi^2}{6}(1+\chi) \right)^2}} \,\,.
\label{VE_chaotic2}
\end{align}
We can distinguish two qualitatively different regimes: For $\chi < -1$, the denominator of Eq.~\eqref{VE_chaotic2} is always strictly positive, whereas for $-1 <  \chi < 0$ the potential features a pole for large values of $\varphi$ (which is however not reached for any finite value of the canonically normalized field $\hat \varphi$). Note that for $\chi=-1$, the transformation to the Jordan frame becomes trivial, $\Omega^2(X= 0) = 1$. Hence in this case $V_E =V_J$ and we reproduce the predictions of standard chaotic inflation with a quadratic potential.  

\begin{figure}
  \centering
\subfigure[Einstein frame scalar potential $V_E$ in terms of the canonical field $\hat{\varphi}$, for  $\chi=-1.01$ and $m=5.6\times10^{-6}$ (the latter fixed by $A_{s}^0$). The vertical dashed black lines represent the field value at the end of inflation and 60-efolds earlier, respectively.]{%
   \includegraphics[width=7cm]{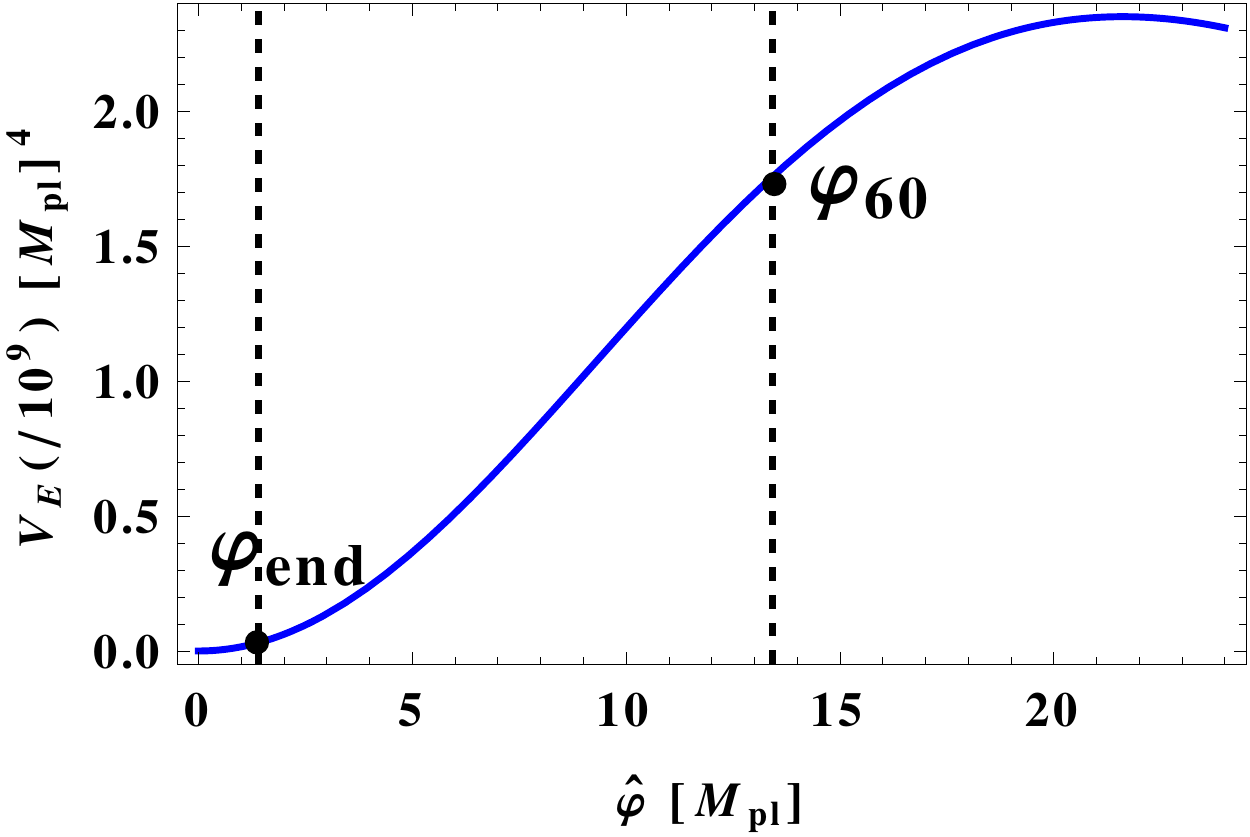}
\label{pot_1}}
~
\subfigure[Inflaton field $\varphi$ vs.\ its canonically normalized counterpart $\hat{\varphi}$ plot for $\chi=-1.01$. Along the inflationary trajectory, $\varphi \approx \hat \varphi$ is a good approximation.]{%
   \includegraphics[width=7.3cm]{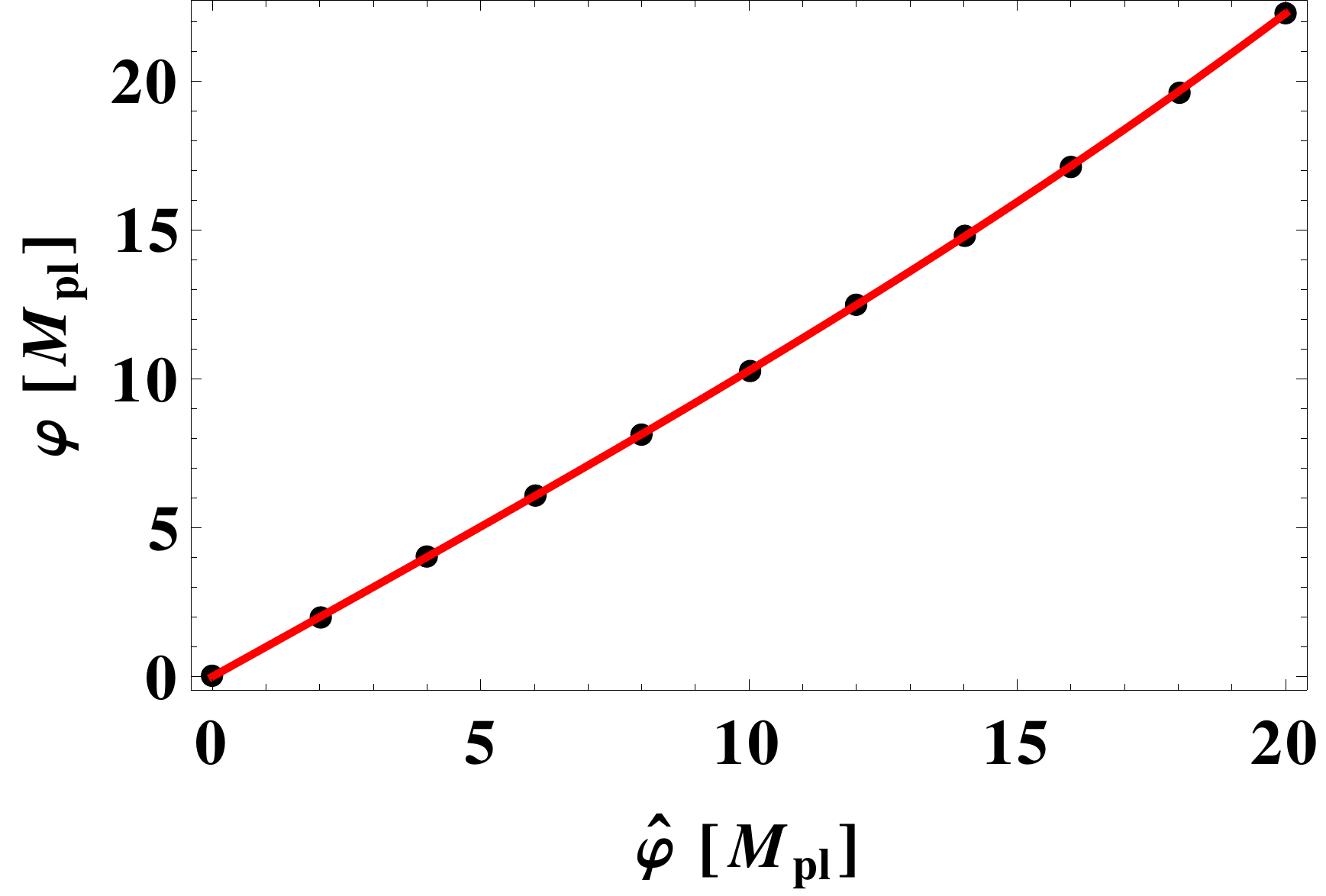} 
\label{phi_phicap}}
\caption{Inflationary dynamics for $W = m \phi X$.}
\end{figure}

We shall first consider the case $\chi < -1$. The inflationary potential in terms of canonical inflation field $\hat \varphi$ is shown in Fig.~\ref{pot_1}. The potential has a minimum at $\varphi = 0$, and it also vanishes for infinitely large field values. Inflation is possible around its maximum when the field rolls towards the minimum at $\varphi = 0$. {Note that this potential has a serious initial value problem for the inflation field. If the initial field value of the inflaton is larger than the position of the maximum of the potential $\varphi_\text{max}$, the field rolls towards the wrong post-inflationary vacua. For our considerations, we assume that the field starts to roll from any field value between $\varphi_\text{max}$ and $\varphi_{60}$ - see Fig.~\ref{pot_1}}. 
The evolution of the inflaton field is governed by the slow-roll equation of motion, 
\begin{equation}
3 K_{\phi \bar \phi} {\cal H} \dot{\varphi} + V_E'( \varphi) = 0 \,,
\end{equation}
where ${\cal H}$ denotes the Hubble parameter during inflation. This can be expressed as
\begin{equation}
\frac{d \varphi}{d N} -  \frac{1}{V_E} \left( \frac{V_E'(\varphi)}{K_{\phi \bar \phi}}\right) = 0 \,,
\label{slow-roll}
\end{equation}
allowing the evaluation of $\varphi(N)$ without explicitly normalizing the inflaton field $\varphi$.

The predictions for the amplitude of scalar perturbations, tilt of the scalar perturbations, and the ratio of tensor and scalar perturbations amplitudes in the Einstein frame are given respectively by
\begin{equation}
A_s = \frac{V_E}{24 \, \pi^2 \, \epsilon_E} \,, \qquad n_s = 1 - 6 \, \epsilon_E + 2 \, \eta_E \,, \qquad r = 16 \, \epsilon_E~,
\end{equation}
evaluated when the CMB pivot scale exited the horizon at $N_* = 60$ e-folds before the end of inflation. The slow-roll parameters $\epsilon_E$ and $\eta_E$ are given by
\begin{equation}
\epsilon_E = \frac{1}{2} \left(\frac{V_E'(\hat \varphi)}{V_E} \right)^2 \,, \quad \eta_E = \frac{V_E''(\hat \varphi)}{V_E} \,,
\end{equation}
where the derivatives with respect to the canonically normalized field $\hat \varphi$ may be evaluated using Eq.~\eqref{eq:derivatives}. For the scalar potential~\eqref{VE_chaotic2}, this leads to
\begin{align}
\epsilon_E &= \frac{2\big(1+\frac{\varphi^2}{6}(1+\chi)\big)^2}{\varphi^2\big(1+\frac{\varphi^2}{6}\chi(1+\chi)\big)}\,,\label{eps1}\\
\eta_E &= \frac{2 (1 + \varphi^2 (1 + \chi) (1 + \frac{1}{6} \varphi^2 (1 + \chi) (\frac{1}{6} + \chi (1 + \frac{1}{18} \varphi^2 (1 + \chi)))))}{\varphi^2(1 + \frac{1}{6}\varphi ^2 \chi  (1 + \chi))^2}\,. \label{eps2}
\end{align}
We note that the slow-roll parameters are functions of $\chi$ and $N$ only, and do not depend on the parameters $m$ and $\zeta$. Note moreover that while the above calculation has been performed in the Einstein frame, identical expressions for $n_s$ and $r$ can be obtained directly in the Jordan frame~\cite{Chiba:2008ia}.

\begin{figure}
  \centering
\subfigure[Predictions for $n_s$ and $r$ for $W=m\phi X$, overlayed with the $68\%$ and $95\%$ CL contours from Planck. The black line represents variations w.r.t $\chi$, where $\chi = -1$ corresponds to the predictions of chaotic inflation. For comparison, the dashed curve shows the predictions of natural inflation.]{
   \includegraphics[height = 6.5cm]{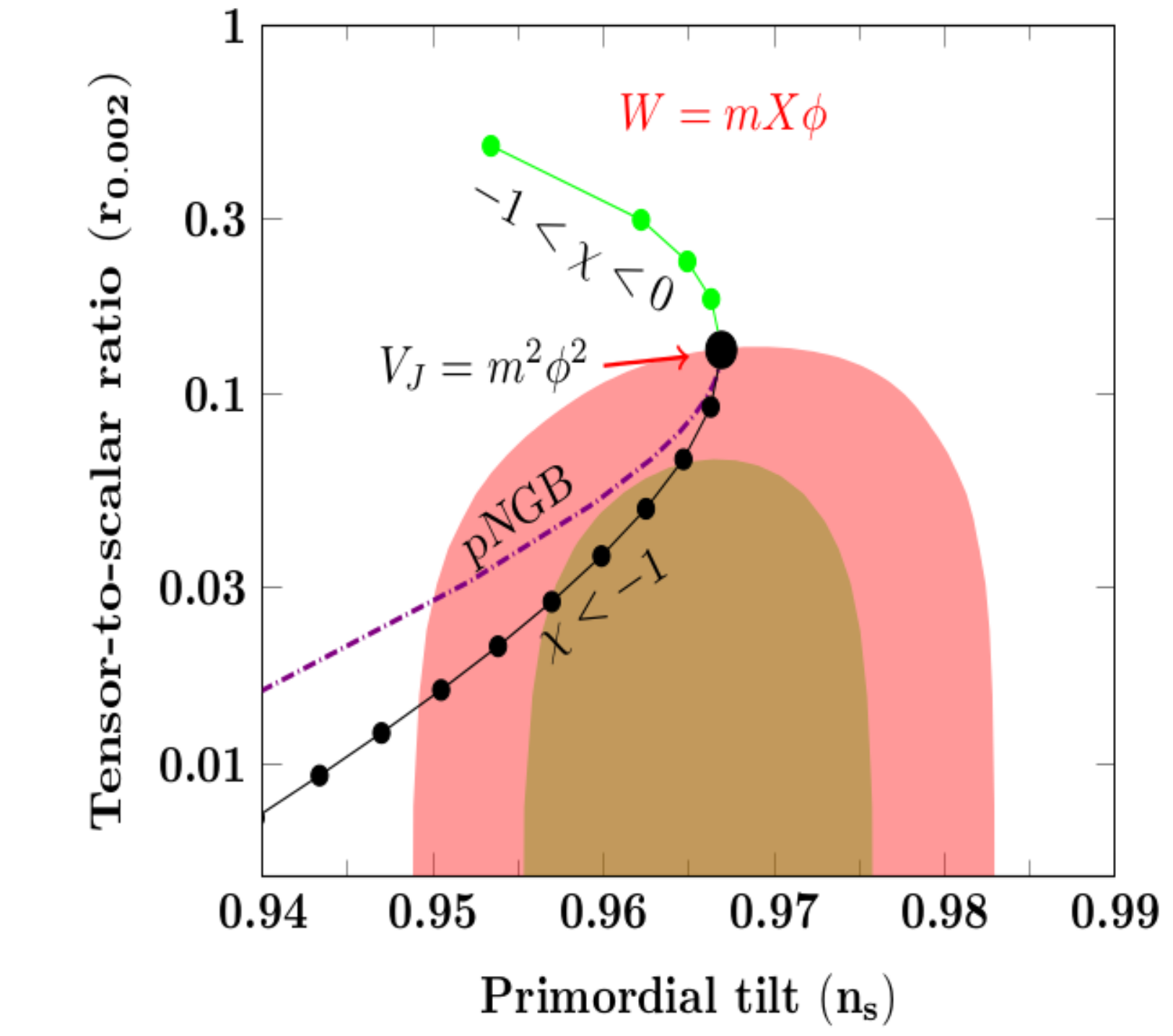}
\label{mphix_nsr}}  \hfill
\subfigure[Amplitude of the scalar perturbations $A_s/A_s^0$ (red dashed line) and its tilt $n_s$ (horizontal contours) as a function of the model parameters $m$ and $\chi$. ]{
   \includegraphics[height = 6.5cm]{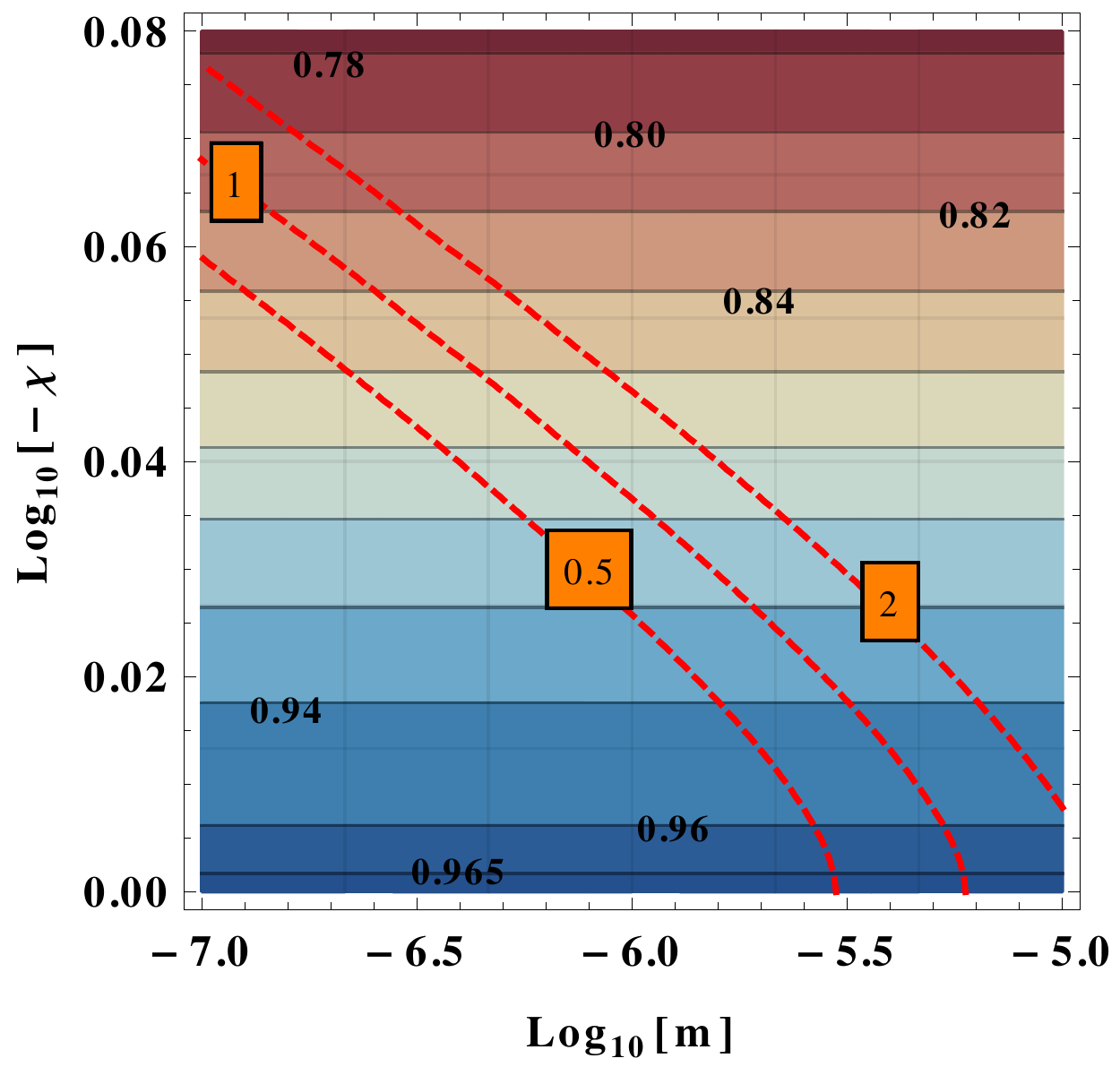}
\label{par_scan}}
\caption{Parameter space and predictions for $W = m \Phi X$.}
\label{fig:predictionsCMB1}
\end{figure}

In Fig.~\ref{fig:predictionsCMB1} we summarize the resulting CMB predictions, obtained by numerically solving the slow-roll equation of motion. Requiring $A_s$ to lie within the $99.7 \%$~CL of the Planck data~\cite{Ade:2015lrj}, we determine the spectral index $n_s$ and the tensor-to-scalar ratio $r$ by numerically solving Eq.~\eqref{slow-roll}. The results are  depicted in Fig.~\ref{mphix_nsr}, with the background contours corresponding to PLANCK 2015 TT + low $l$ polarization~\cite{Ade:2015lrj}. The black dots indicate different values of $\chi$ parameter with $\chi$ varying from $-1$ to about $-1.03$ with decreasing $r$ and $n_s$.  For $\chi < -1.03$, the spectral index $n_s$ lies well outside the Planck contour, see also Fig.~\ref{par_scan} which shows the dependence of $n_s$ on the pair of parameters $\{m,\chi\}$, together with the curvature perturbation $A_s$ (dashed red lines) normalized to the Planck best-fit value $A_{s}^0$: For increasing values of $|\chi|$, the spectral index decreases until it reaches values well beyond the current observational bound. Some of these contour lines are marked in the figure.

In the vicinity of the limiting case of chaotic inflation ($\chi = -1$), the CMB predictions may be understood analytically. For $\delta \equiv \frac{\varphi^2}{6}\chi(1+\chi) \ll 1$ and $\varphi(N) \gg \varphi_\text{end}$,
integrating Eq.~\eqref{slow-roll} yields
\begin{equation}
4 N \simeq \varphi(N)^2 \frac{(1 - \chi)}{2} + \frac{(\varphi^2(\chi -1) + 12)\delta}{4 \chi} +  \frac{(\varphi^2(1 -\chi) + 9)\delta^2}{6 \chi^2} + \mathcal {O} (\delta^3)  \,.
\label{phi_in_efold}
\end{equation}
 In the limit $\chi \rightarrow -1$, we recover the familiar expressions of quadratic chaotic inflation,
\begin{equation}
\varphi^2 \simeq 4 N \,, \qquad \epsilon_E \simeq \frac{1}{2 N} \,,  \qquad \eta_E \simeq \frac{1}{2 N} \,.
\end{equation}
The $\chi$-dependence in the vicinity of $\chi \rightarrow -1$ limit can be understood by expanding the spectral index and tensor-to-scalar ratio in terms of the expansion parameter $\delta$. Employing Eq.~\eqref{phi_in_efold}, we obtain  
\begin{align}
n_s &\simeq 1 - \frac 2 N + \frac {2} {27} (\chi + 1)^2 (3\chi^2 +2\chi-14)N +\mathcal{O}(\delta^3{/N})  \,,  \label{eq:nexp} \\
r &\simeq\frac 8 N  - \frac {8}{3} (\chi + 1)(\chi-4) + \mathcal{O}(\delta^2{/N})  \,,\label{eq:rexp} 
\end{align} 
{where $\delta \simeq 2/3 N \chi (1 + \chi)$.}
Qualitatively the $\chi$-dependence in the $n_s - r$ plane as shown in Fig.~\ref{mphix_nsr} is similar to the result found for the natural inflation (pNGB inflation) potential $V \sim \big(1-\text{cos}\frac{\varphi}{f}\big)$~\cite{Freese:1990rb}. In the latter case, expanding in powers of $1/f \ll 1$, with $f$ being the axion decay constant in Planck units, yields
\begin{align}
n_s &\simeq 1 - \frac 2 N - \frac{N}{6 f^4} +\mathcal{O}(1/f^8)  \,,   \label{eq:nexp_nat} \\
r &\simeq\frac 8 N - \frac{4}{f^2} + \mathcal{O}(1/f^4) \,.
\label{eq:rexp_nat}
\end{align}
For a fixed value of $n_s$, comparing Eqs.~\eqref{eq:nexp} and \eqref{eq:nexp_nat} we find $1/f^4 = 4/9(\chi + 1)^2(3\chi^2 +2\chi-14)$.  We then immediately see that the value of $r$ from Eq.~\eqref{eq:rexp} is  in fact smaller than the natural inflation counterpart from Eq.~\eqref{eq:rexp_nat}. Hence the two models yield similar, but not identical predictions in the $n_s -r$ plane. As the numerical analysis shows, this is true in the entire parameter range of interest, see Fig.~\ref{mphix_nsr}.

Finally we turn our attention to the case $-1 < \chi < 0$. Again we can identify the real axis in the complex plane spanned by $\phi = (\varphi + i \tau)/\sqrt{2}$ as the only viable inflationary trajectory (the trajectory along the imaginary axis exhibits a tachyonic instability).  We can achieve enough efolds of slow-roll inflation only if the parameter  $\chi$ is not too different from $\chi = -1$ (otherwise the potential becomes too steep). However even in this case, the predicted tensor-to-scalar ratio becomes larger than the usual quadratic chaotic inflation model, and remains outside the $2$-$\sigma$ contour of PLANCK data - see Fig.~\ref{mphix_nsr}. 

The example discussed above explicitly demonstrates some of the effects discussed in Sec.~\ref{sec:attractors}. In the regime $\chi < -1$ (corresponding to $\xi > 0$), we find ourselves in the case (i) of Sec.~\ref{sec:attractors}. In terms of the canonically normalized field, the scalar potential in the Einstein frame vanishes for large field values. Together with $V \propto \varphi^2$ at small field values, this leads to a hilltop-type potential. {Similarly for the case $0 > \chi > -1$ (i.e $\xi < 0$) the potential is exponentially steep and the model produces a too large tensor-to-scalar ratio (far away from the attractor point in the $n_s$-$r$ plane).}


\subsection{Starobinsky inflation from \texorpdfstring{$W=\lambda \phi^2 X$}{W=lambda phi2 X}} \label{phi2x}

We now turn to an example of the case (ii) of Sec.~\ref{sec:attractors}: $W=\lambda \phi^2 X$, see also Ref.~\cite{Lee:2010hj}. The Jordan frame scalar potential following Eq.~\eqref{potn_jr} is $V_J = \lambda^2 \phi^4 $. As in the previous section, we can restrict ourselves without loss of generality to $\chi < 0$, and as above we find that both $X$ and $\hat \tau$ settle to zero vevs due to a large masses compared to the Hubble scale during inflation. The $F$-term scalar potential along the inflationary direction in the Einstein frame is given by
\begin{align}
V_E =\dsl{\frac{\lambda^2\varphi^4}{4\left(1-\frac{\varphi^2}{6}(1+\chi)\right)^2}} \; ,
\label{lphix_potn}
\end{align}
where we have used $\varphi = \sqrt{2}$\,Re\,$\{\phi\}$. 
In contrast to the previous example, the above potential asymptotically approaches a constant value, due to the quartic power of the inflaton field in the numerator. 
\begin{figure}[t]
    \centering
    \includegraphics[width=7.0cm]{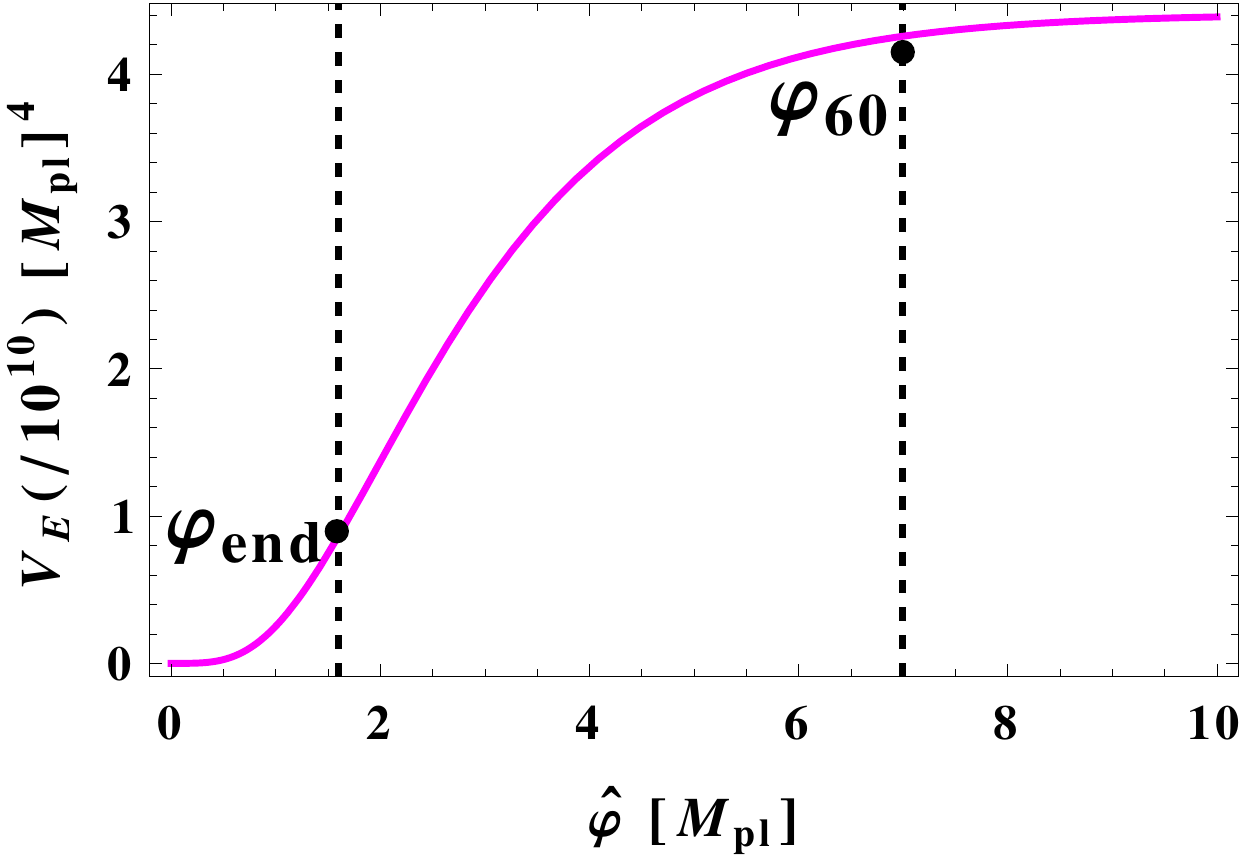} 
             \caption{Plot of $V_E$ in terms of the canonical field $\hat{\varphi}$ for  $W = \lambda \phi^2 X$. The parameter values are $\chi=-3$ and $\lambda=1.4\times10^{-5}$ 
(fixed by $A_s^0$). The dashed vertical lines represent the field value at the end of inflation and 60-efolds earlier, respectively.} 
\label{lphix_pot}
\end{figure} 

The potential in the limit of  $\chi=-1$ is a simple $\varphi^4$ potential, {and in this case the kinetic term becomes canonical with the conformal factor being unity}. {Therefore, the potential in the Einstein frame is too steep,} and is disfavoured by the PLANCK data \cite{Ade:2015lrj}. 
We next turn to the case $\chi < -1$. A plot of this potential after canonical normalization of the kinetic term of $\varphi$ is shown in Fig.~\ref{lphix_pot}, where $\hat{\varphi}$ is the canonical inflaton field. It is also a two parameter $\{\lambda,\chi\}$ potential. For a given $\chi$ and for large field values this potential has a long plateau type region with a slowly varying slope. 
\begin{figure}[t]
  \centering
\subfigure[$n_s$ - $r$ plot for $W=\lambda\phi^2 X$ overlayed with the $68\%$ and $95\%$ CL contours from 2015 Planck data. For $|\chi| \gg1$ predictions are asymptotically close to Starobinsky inflation model shown by blue triangle. ] {
   \includegraphics[height = 6.7cm]{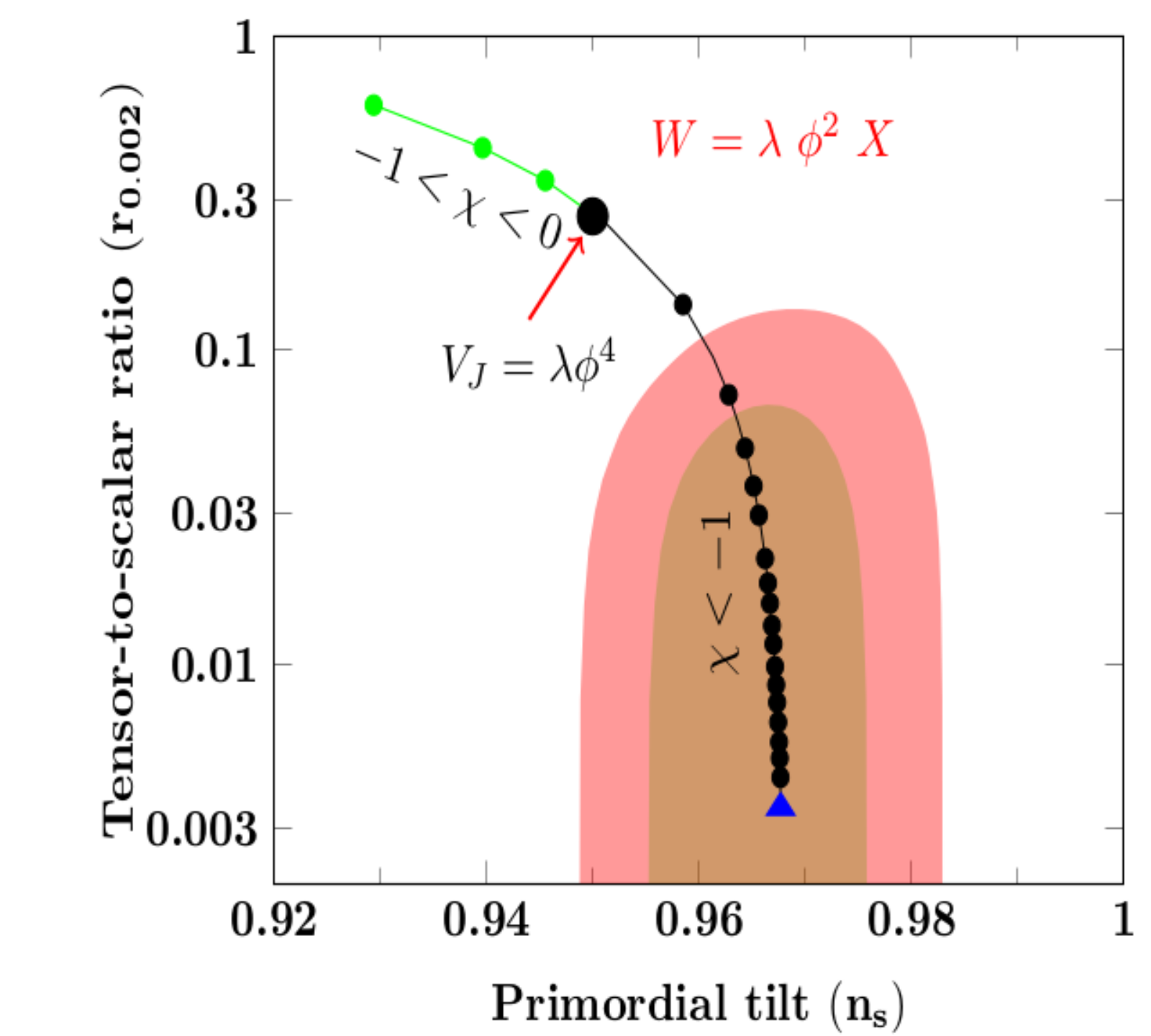}
\label{quartic_nsr}}  \hfill
\subfigure[Amplitude of the scalar perturbations $A_s$ (red dashed line) and its tilt $n_s$ (vertical contours) as a function of the model parameters $\lambda$ and $\chi$. ]{%
   \includegraphics[height = 6.7cm]{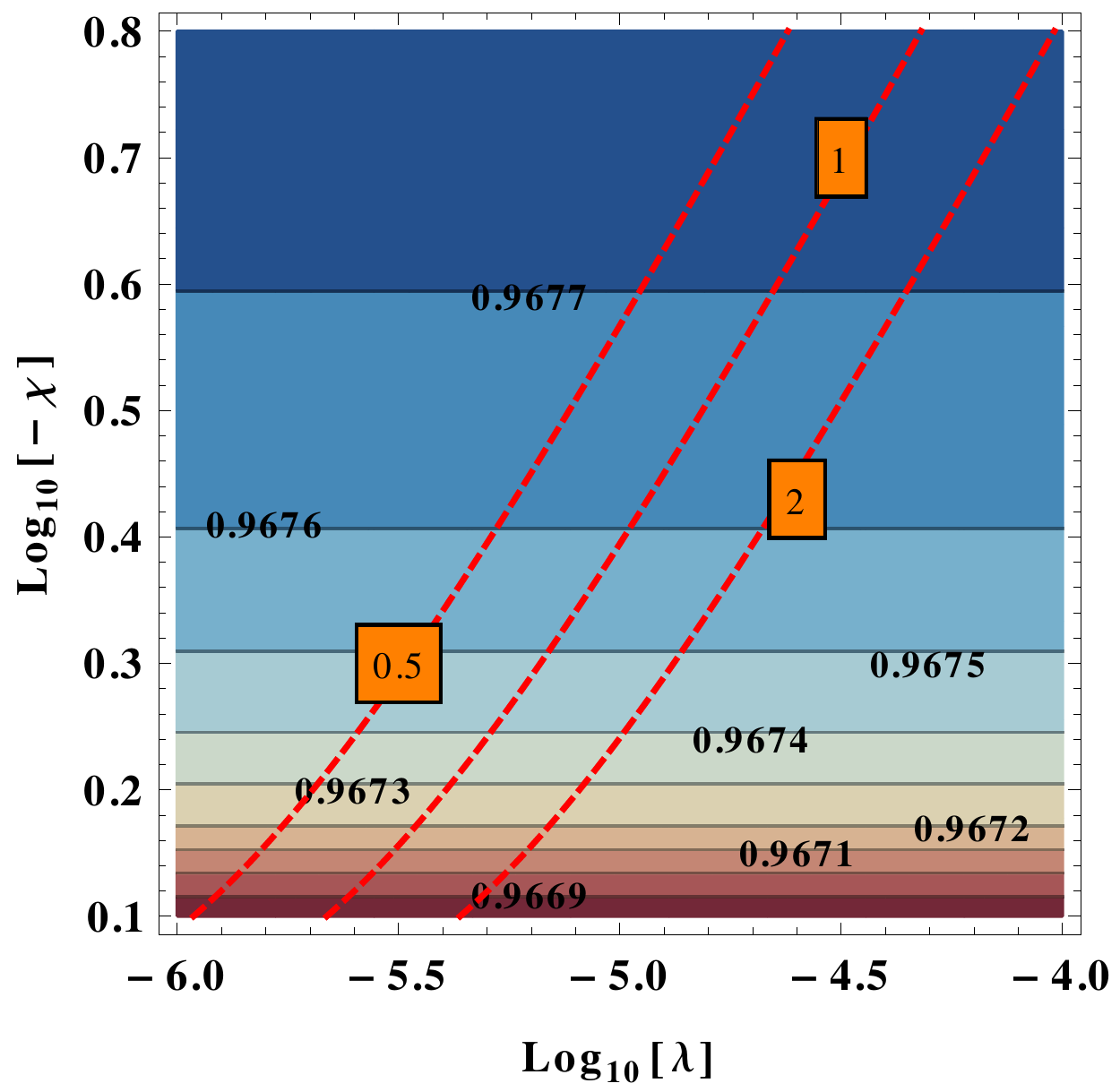}
\label{quartic_parscan}}
\caption{Parameter space and predictions for $W = \lambda \Phi^2 X$.}
\end{figure}
 In fact, this potential can accommodate successful inflation for wide range of values of the $\chi$ parameter. Increasing $\chi$ will add more flatness to the potential. The value of $\lambda$ will be fixed by the amplitude of density perturbations. To demonstrate this feature we show the variation of $\chi$ from  $-1$ to $-5$ (from top to bottom) in the usual $n_s$ vs. $r$ plot, cf.\ Fig.~\ref{quartic_nsr}.
In the limit $\chi\to -\infty$ the predictions asymptotically approach to those of the Starobinsky inflation model \cite{Starobinsky:1980te} (see next paragraph). Fig.~\ref{quartic_parscan} shows the variation of $n_s$ as well as the normalized curvature perturbation $A_s/A_{s}^{0}$ (red dashed lines) with respect to the model parameters $\{\lambda,\chi\}$. Again $n_s$ is independent of model parameter $\lambda$. The asymptotic behaviour of the spectral index as $|\chi|$ increases can be clearly seen in this plot. 

This behaviour can be well understood by considering the following analytic expressions of inflationary slow-roll parameters given by
\begin{align}
\epsilon_E &=  \frac{8}{\varphi^2 (1 + \frac{1}{6} \varphi^2 \chi (1 + \chi)) } \,,\\
\eta_E &= \frac{12 + \frac{1}{6} \varphi^2 (1 + 3 \chi + 2 \chi^2) +  \frac{2}{9} \varphi^4 \chi (1 + \chi)^2}{\varphi^2 (1 + \frac{1}{6} \varphi^2 \chi (1 + \chi))^2}\,,
\end{align}
with
\begin{equation} 
8  N = - \varphi^2 \chi - 6 \ln\big(1 - \frac{1}{6} \varphi^2(1 + \chi )\big) \, ,
\label{efold_eq}
\end{equation}
where Eq.~\eqref{efold_eq} is obtained by integrating the slow-roll Eq.~\eqref{slow-roll} with the potential given by Eq.~\eqref{lphix_potn}. For $\varphi^2 (1 + \chi)/6 \ll 1$, this yields $8 N \simeq \varphi^2$ whereas for $\varphi^2(1 + \chi)/6 \gg 1$ we find $8 N \simeq - \varphi^2 \chi$. 
So in the former case, we find
\begin{equation}
\epsilon_E \simeq \frac{1}{N} \,, \quad \eta_E \simeq \frac{3}{2 N} \quad \rightarrow n_s \simeq 1 - \frac{3}{N} \,, \quad r \simeq \frac{16}{N}~,
\end{equation}
which are just the results for chaotic inflation with a quartic potential. In the latter case, we find
\begin{equation}
\epsilon_E \simeq \frac{3 \alpha}{4 N^2} \,, \quad \eta_E \simeq - \frac{1}{N} \quad \rightarrow n_s \simeq 1-\frac 2 N \,, \quad r\simeq \frac{12 \alpha }{N^2} \,,
\label{nsr_quartic}
\end{equation}
with $\alpha = \chi/(1 + \chi)$, which are the predictions of the so-called $\alpha$-attractors~\cite{Kallosh:2013hoa}. For $\alpha = 1$ ($|\chi| \rightarrow \infty$) we obtain the predictions of the Starobinsky model. We thus explicitly see the mechanism described in case (ii) of Sec.~\ref{sec:attractors} at work here.

Finally, we consider the case $ -1<\chi < 0$ i.e $\xi < 0$. As in the example of Sec.~\ref{mphix}, the $\varphi$ direction is identified as the only possible inflationary trajectory, with the orthogonal direction stabilized during inflation. The prediction for tensor-to-scalar ratio exceeds the one for quartic inflation with a canonical kinetic term - see Fig.~\ref{quartic_nsr}. As this is well outside the $2$-$\sigma$ contour of PLANCK data, we can conclude that $-1<\chi < 0$ is not a viable parameter range for this model.
This is in agreement with the general argument of Sec.~\ref{sec:attractors}.


\section{Conclusions \& Outlook} \label{conclusion}

Current CMB data allow for an energy scale of inflation as high as about $10^{16}$~GeV. At these energies, supergravity effects can no longer be neglected. They may spoil the flatness of the inflationary direction, destabilize the inflationary trajectory (as e.g.\ in F-term hybrid inflation~\cite{Ferrara:2010yw,Ferrara:2010in,Lee:2010hj}), or even also improve the flatness of the potential (as in models with non-minimal coupling to gravity such as Higgs inflation~\cite{Bezrukov:2007ep} or $\alpha$-attractors~\cite{Kallosh:2013yoa}). In this paper we have systematically studied supergravity contributions to Jordan frame inflation models, i.e.\ inflation models characterized by a non-minimal coupling to gravity and canonical kinetic terms in the Jordan frame. Our focus here is on single-field inflation models driven by F-term potentials, for an example of D-term inflation in this setup see~\cite{Buchmuller:2012ex}.

We disentangle two types of supergravity contributions in the Jordan frame, arising from contributions to the scalar potential and from the non-minimal coupling to gravity.
We find that the former generically yields a contribution to the Jordan frame scalar potential which is at most of the same power in the inflaton field as  the contribution from global supersymmetry. 
We moreover derive the condition on the superpotential for which this term vanishes identically (cf.\ Eq.~\eqref{eq:condition}) and find that for single-field inflation models this corresponds to the condition of a vanishing superpotential during inflation.

In a second step, we turn to the effects of the non-minimal coupling to gravity, which translate to non-canonical kinetic terms in the Einstein frame. As observed e.g.\ in the context of $\alpha$-attractors~\cite{Kallosh:2013yoa}, this can lead to an exponential flattening of the scalar potential in the Einstein frame in terms of the canonically normalized field. However, this mechanism requires the powers of the inflaton field appearing in the non-minimal coupling to gravity and in the (Jordan frame) scalar potential to be adjusted accordingly. 

The findings of this paper are illustrated in various examples. In particular we focus on two examples of the type $W = \lambda X \phi^n$, with $n = \{1,2\}$. The CMB data can be reproduced for certain values of the parameter $\chi$, which parametrizes the non-minimal coupling to gravity. In particular in the $n=2$ case we asymptotically reproduce the Starobinsky inflation model for $|\chi| \gg 1$. Since in all these models the superpotential vanishes along the inflationary trajectory, the supergravity contributions to the Jordan frame potential are identically zero. We further comment on tribrid inflation models, which, in contrary to the more commonly discussed hybrid inflation models, are protected from supergravity contributions to the Jordan frame scalar potential. However, this protection does not encompass the second dynamical degree of freedom in these models, the waterfall field. Using two different realizations of tribrid inflation we illustrate how this may lead to a destabilization of the inflationary trajectory or to a potentially viable inflation model.

Our results may be used as guidelines to easily estimate the effect of supergravity contributions in a given Jordan frame inflation model. They may moreover be useful in inflationary model building to easily understand what type of terms in the superpotential or K\"ahler potential may help modify a given scalar potential in a desired way through supergravity contributions.

In this paper, we focused on a simple frame function motivated by approximate scale invariance and on superpotentials which can be expressed as polynomials in the inflaton field. It would be interesting to extend this work beyond these two assumptions. Moreover, the models studied in this paper should be considered as illustrative toy models, at this point without a deeper motivation from particle physics and also lacking a study of the subsequent cosmology after inflation. In particular, we have not addressed the question of (low-energy) supersymmetry breaking.


\section*{Acknowledgements}
K. Das is supported by the DAE fellowship from Saha Institute of Nuclear Physics. V.D.\ acknowledges the financial support of the UnivEarthS Labex program at Sorbonne Paris Cit\'e (ANR-10-LABX-0023 and ANR-11-IDEX-0005-02), the Paris Centre for Cosmological Physics and the l'Or\'eal-Unesco program. K. Dutta would like to thank ICTP, Trieste via its Junior Associateship programme for hospitality when the work was initiated and partially done, and Max Planck Institute for Physics, Munich where a part of the work was completed. KD is partially supported by a Ramanujan Fellowship and a Max Planck Society-DST Visiting Fellowship.  
\appendix
\section{Tribrid inflation in CSS}
\label{appendixa}

In this Appendix, we will explore the supergravity effects for tribrid inflation models described by the following superpotential \cite{Antusch:2008pn}
\begin{align}
W= \kappa X (H^l - M^2) + \lambda\phi^n H^m \,,
\end{align}  
in the context of Jordan frame supergravity.
In this kind of models, inflation ends via a phase transition that is triggered by the mass of the waterfall field $H$ becoming tachyonic as the inflaton field rolls towards smaller field values. 

As in the main text, we will assume the following frame function
\begin{align} \label{frame_appendix}
\Phi = -3 + |\phi|^2 + |X|^2 + |H|^2 - \zeta|X|^4 + \frac{\chi}{2}(\phi^2 + \bar{\phi}^2)~,
\end{align}
and we will restrict the superpotential by our choice to $(n,m)\le2$ and $l=2$. Constraints on the powers $l,m$ and $n$ have been discussed in the literature \cite{Antusch:2012bp,Antusch:2012jc}. We would like to emphasize that the tribrid inflation models satisfy the conditions of Eq.~\eqref{impt_condt} of vanishing supergravity corrections along the inflationary trajectory in the Jordan frame. But the supergravity corrections to the fields orthogonal to the inflaton directions (e.g waterfall $H$ field) are not protected from these corrections, and are potentially dangerous in spoiling the waterfall mechanism.

We start with the case $n = m = 2$, see also~\cite{Antusch:2009ef}.
In the globally supersymmetric limit, this model leads to a tree-level flat potential, lifted by one-loop corrections. The spectral index is found to be $n_s \simeq 0.98$ and may be lowered by supegravity contributions~\cite{Antusch:2009ef}. These results are reproduced here for $\chi = -1$ (after taking into account the effective one-loop potential arising after integrating out the waterfall fields). Departing from $\chi = -1$, the supergravity contributions can result in a positive ($\chi > -1$) or negative ($\chi < -1$) slope of the tree-level potential (this is just the effect of the non-canonical kinetic term discussed in Sec.~\ref{sec:attractors}). For values of $\chi$ sufficiently close to $\chi = -1$, these small corrections may modify the globally supersymmetric predictions in an interesting way. We leave a detailed investigation to future work. 

Next, we consider the case $n = 2$, $m = 1$. In this case, the globally supersymmetric tree-level scalar potential for the inflaton $\varphi = \sqrt{2}\,$Re$\,\{\phi\}$ is given by 
\begin{align} 
V^\text{glob} = V_{J} = \kappa^2 M^4 + \frac{\lambda^2 \varphi^4}{4}~,
\label{tribrid_tree}
\end{align}
where we have assumed  $\langle X \rangle=0$ and $\langle H \rangle = M$ during inflation\footnote{It can be shown that the canonically normalized imaginary part of the $\phi$ field has a mass larger than the Hubble scale during inflation, and it settles to zero vev.}. In global supersymmetry, this potential is simply too steep to yield a viable inflation model in agreement with the current data. However, given our discussion in Sec.~\ref{sec:attractors}, we may hope to achieve a sufficient flattening of the scalar potential when transforming to the Einstein frame taking into account the canonical normalization of the inflaton field in the Einstein frame.

Switching to the Einstein frame, the complete potential including the waterfall fields becomes,
\begin{align}
V_{E}(\varphi,H) = \frac{ \big((H^2 - M^2)(\bar{H}^2 - M^2)\kappa^2 + \frac{\lambda^2\varphi^4}{4}\big)\big(1+\frac{\varphi^2}{6}\chi(1+\chi)\big) + 2|H|^2\lambda^2\varphi^2(1+\frac{\varphi^2}{6}\chi) 
}  
{\big(1+\frac{\varphi^2}{6}\chi(1+\chi)\big)\big(1-\frac{|H|^2}{3}-\frac{\varphi^2}{6}(1+\chi)\big)^2}.   
\end{align}
Note that the $H$ field is non-canonical in the Einstein frame ($\because K_{H\bar{H}}\ne1$), and the masses of the canonical waterfall fields in the Einstein frame during inflation are given by 
\begin{align} 
m_{\hat H_R}^2 &= \frac{- 2\kappa^2 M^2 + 2 \lambda^2 \varphi^2}{(1 - \frac{\varphi^2}{6}(1 + \chi))^2}\\ &+ \frac{ 2\kappa^2 M^4}{3(1\! -\! \frac{\varphi^2}{6}(1\! +\! \chi))^2} \! + \frac{ \lambda^2 \varphi^4}{6(1 - \frac{\varphi^2}{6}(1 \!+\! \chi)\!)^2}\! \left(\!1 + \frac{2\chi(1\! - \!\frac{\varphi^2}{6}(1+ \chi)\!)(1\! -\! \frac{\kappa^2 M^2}{\lambda^2 \varphi^2}(1+ \chi)\!)}{(1 + \frac{\varphi^2}{6}(1 + \chi))^2} \!\right) ~, \nonumber
\end{align}

\begin{align} 
m_{\hat H_I}^2 &= \frac{ 2\kappa^2 M^2 + 2 \lambda^2 \varphi^2}{(1 - \frac{\varphi^2}{6}(1 + \chi))^2}\\ &+ \frac{ 2\kappa^2 M^4}{3(1\! -\! \frac{\varphi^2}{6}(1\! +\! \chi))^2} \! + \frac{ \lambda^2 \varphi^4}{6(1 - \frac{\varphi^2}{6}(1 \!+\! \chi)\!)^2}\! \left(\!1 + \frac{2\chi(1\! - \!\frac{\varphi^2}{6}(1+ \chi)\!)(1\! +\! \frac{\kappa^2 M^2}{\lambda^2 \varphi^2}(1+ \chi)\!)}{(1 + \frac{\varphi^2}{6}(1 + \chi))^2} \!\right) ~. \nonumber
\end{align}

In the limit of $\chi \rightarrow  -1$,
\begin{align} 
m_{\hat H_I, \hat H_R}^2  = \pm 2 \kappa^2 M^2 + 2\lambda^2 \varphi^2 + \frac{2}{3} \kappa^2 M^4 - \frac{1}{6}\lambda^2 \varphi^4~,
\end{align}
where the first two terms are the global SUSY mass terms for the waterfall fields. The last two terms provide the supergravity corrections suppressed by $M_{pl}^2$. In particular the last term in the above expression makes the waterfall mass tachyonic for large inflaton field values $\varphi \gsim M_{pl}$. On the other hand, the tachyonic instability for small field values, at $\varphi^2 < \varphi_c^2 \leq \kappa^2 M^2/ \lambda^2$ indicates the usual waterfall instability which ends inflation. Thus the requirement of a viable waterfall mechanism limits the inflaton field range. As we move away from $\chi \simeq -1$, the field range for which the mass for the waterfall fields remain positive shrinks further, introducing the necessity to fine-tune the initial value of the waterfall field, until finally it even becomes impossible to account for 60 e-folds of inflation. 

However, in order to achieve a sufficient flattening of the scalar potential in the Einstein frame, as discussed in Sec.~\ref{sec:attractors}, we need sufficiently large values of $\phi^2 |1 - \chi|$. In this sense, the observed destabilization of the waterfall field at large field values prevents us from constructing a viable inflation model.

One might hope to resolve the problem of the tachyonic instability in the waterfall field mass at large inflaton-values by adding a term $-\zeta_H |H|^4$ term in the frame function of Eq.~\eqref{frame_appendix}, as for the stabilizer field $X$. However, this will also prevent the desired destabilization of the waterfall field at $\varphi < \varphi_c$, an essential feature of tribrid inflation.

We point out that on the contrary, in the $n = m = 2$ case discussed above, the 
the waterfall field is not destabilized at large inflaton values. This can be understood by looking at the second line of Eq.~\eqref{impt_condt}, which yields a negative mass term for $H$ for $n = 2$, $ m = 1$ but not in the case $n = m = 2$.

In all tribrid models, once the auxiliary field is stabilized, it has nothing to do with the field dynamics. Now as the critical point is approached where the waterfall transition has to take place, the non-trivial dynamics is governed by two fields simultaneously. So in the two-dimensional $(\phi,H)$-space the requirement of $V_{\Delta SUGRA}$ to be zero in $\phi$-direction ($i.e$ satisfying the conditions of Eq.~\eqref{impt_condt}) alone is insufficient to comply with the dynamics.  We recall that the usual hybrid inflation model given by Eq.~\eqref{hybrid} is also difficult to implement in this framework as the imaginary component of the complex inflaton field has a tachyonic mass \cite{Buchmuller:2012ex}.


\end{document}